\documentclass[11pt]{article}
\usepackage[final]{acl}

\usepackage{times}
\usepackage{latexsym}
\usepackage{multirow}
\usepackage{caption}
\usepackage[T1]{fontenc}
\usepackage{booktabs}

\usepackage{microtype}
\usepackage{amsmath}

\usepackage{inconsolata}

\usepackage{graphicx}
\usepackage{float}
\usepackage{subcaption}

\title{polyBART: A Chemical Linguist for Polymer Property Prediction and Generative Design}

\author{
 \textbf{Anagha Savit\textsuperscript{1}},
 \textbf{Harikrishna Sahu\textsuperscript{1}},
 \textbf{Shivank Shukla\textsuperscript{1}},
 \\
 \textbf{Wei Xiong\textsuperscript{1}},
 \textbf{Rampi Ramprasad\textsuperscript{1}}
\\
 \textsuperscript{1} School of Materials Science and Engineering, Georgia Institute of Technology
\\
 \small{
   \textbf{Correspondence:} \href{}{rampi.ramprasad@mse.gatech.edu}
 }
}

\begin{document}
\maketitle
\begin{abstract}

Designing polymers for targeted applications and accurately predicting their properties is a key challenge in materials science owing to the vast and complex polymer chemical space. While molecular language models have proven effective in solving analogous problems for molecular discovery, similar advancements for polymers are limited. To address this gap, we propose polyBART, a language model-driven polymer discovery capability that enables rapid and accurate exploration of the polymer design space. Central to our approach is Pseudo-polymer SELFIES (PSELFIES), a novel representation that allows for the transfer of molecular language models to the polymer space. polyBART is, to the best of our knowledge, the first language model capable of bidirectional translation between polymer structures and properties, achieving state-of-the-art results in property prediction and design of novel polymers for electrostatic energy storage. Further, polyBART is validated through a combination of both computational and laboratory experiments. We report what we believe is the first successful synthesis and validation of a polymer designed by a language model, predicted to exhibit high thermal degradation temperature and confirmed by our laboratory measurements. Our work presents a generalizable strategy for adapting molecular language models to the polymer space and introduces a polymer foundation model, advancing generative polymer design that may be adapted for a variety of applications.

\end{abstract}

\section{Introduction}

Polymers play a crucial role in our daily lives, serving as essential constituents of countless materials and products that we rely on. Designing novel application-specific polymers, however, continues to pose a significant challenge owing to the vast polymer chemical space. The application of Machine Learning (ML) in this space (\citealp{10.1063/1.5023563}; \citealp{Batra2021Emerging}; \citealp{Tran2024}) has made significant strides in addressing the forward problem of predicting material properties from polymer structures (\citealp{10.1063/5.0023759}; \citealp{doi:10.1021/acsami.2c08301}). In contrast, progress on the inverse problem, namely, rapidly designing polymers that meet target property requirements, has been far more limited, largely due to the inherent difficulty of generating chemically valid polymer structures with traditional ML.

Language models, owing to their generative capabilities, have emerged as an effective strategy for addressing both forward and inverse problems in the molecular space. Drawing inspiration from foundation models in Natural Language Processing (NLP), molecular foundation models (\citealp{edwards2022translationmoleculesnaturallanguage}; \citealp{doi:10.1021/acs.jcim.1c00600}) are pretrained on the 'language' of molecules, typically represented as SMILES (\citealp{doi:10.1021/ci00057a005}), SMARTS (\citealp{DaylightSMARTS2007}), or SELFIES (\citealp{Krenn_2020}) strings. These models are subsequently fine-tuned on downstream tasks, such as property prediction, hence solving the forward problem. The problem of generative design is often achieved by navigating the latent space learned during pretraining. The achievements of language models for molecules highlight a key opportunity for extending similar methodologies to polymers. Particularly, the structural and representational parallels between the molecular and polymer domains present a unique possibility of adapting existing molecular foundation models to the polymer space.

Recognizing this, we develop polyBART by strategically leveraging the chemical priors learned by existing molecular foundation models. To accomplish this, we introduce a novel representation, Pseudo-polymer SELFIES (PSELFIES), an extension of SELFIES (\citealp{Krenn_2020}) adapted for polymers. SELFIES (Self-Referencing Embedded Strings) is a robust molecular string representation that guarantees 100\% syntactic validity, meaning every SELFIES string corresponds to a valid molecule. From a chemical standpoint, PSELFIES allow us to represent polymers in a format that mirrors molecular syntax, enabling direct compatibility with existing molecular language models and facilitating seamless continued pretraining. Further details on this representation is found in Section~\ref{sec:pselfies}.

We develop polyBART through the continued pretraining of SELFIES-TED (\citealp{selfiesTED}), an encoder-decoder model based on BART (\citealp{lewis2019bartdenoisingsequencetosequencepretraining}) and designed for molecular representations. PolyBART is a unique unifying model capable of solving both forward and inverse problems in polymer informatics. Our work makes the following key contributions:
\begin{itemize} 
    \item We introduce PSELFIES and develop polyBART, a polymer foundation model that is, to our knowledge, the first application of language models to generative polymer design.
    \item We provide comprehensive computational validation of polyBART, demonstrating strong performance in both polymer property prediction and generation.
    \item We report laboratory synthesis and testing of a polymer predicted by polyBART, which, to our knowledge, is the first successfully validated polymer design guided by a language model. Notably, upon thermal property testing, we find that the experimental measurements align with the model's predictions.
\end{itemize}
\section{Related Work}
\subsection{Chemical Foundation Models}

Recent work has successfully adapted transformer architectures (\citealp{NIPS2017_3f5ee243}) to the molecular domain, achieving impressive results (\citealp{Ross2022}). Early work has focused on adapting well-established transformer backbones, originally developed for natural language understanding, to molecule tasks by training them on molecular string representations. BERT (\citealp{devlin2019bertpretrainingdeepbidirectional}) has been widely used for this purpose, forming the foundation for models like SMILES-BERT (\citealp{10.1145/3307339.3342186}) and Mol-BERT (\citealp{https://doi.org/10.1155/2021/7181815}). Similarly, ChemBERTa (\citealp{chithrananda2020chembertalargescaleselfsupervisedpretraining}) builds on the RoBERTa architecture (\citealp{liu2019robertarobustlyoptimizedbert}) and is pretrained on millions of SMILES strings. Later works have explored the application of Large Language Models (LLMs) to these tasks. Unlike the earlier transformer-based methods, LLMs benefit from their exposure to vast amounts of data, eliminating the need for the pretraining step. For instance, \citet{gruver2024finetunedlanguagemodelsgenerate} demonstrated that instruction fine-tuned LLaMA models (\citealp{touvron2023llamaopenefficientfoundation}) can effectively generate crystal structures. Similarly, \citet{jablonka2024leveraging} showed that GPT-3 (\citealp{NEURIPS2020_1457c0d6}) can be fine-tuned using natural language to perform a wide range of tasks in chemistry and materials science.

Alongside these developments, other works have shown that transformer models can also be used to generate molecular structures and not just to predict their properties. Models like MolT5 (\citealp{edwards2022translationmoleculesnaturallanguage}) and MolGPT (\citealp{doi:10.1021/acs.jcim.1c00600}) enable de novo molecule generation, MolT5 from textual descriptions, and MolGPT through next-token prediction. The SELFIES-TED model generates molecules by exploring its learned latent space. Diffusion methods have also emerged as a strong candidate for molecular generation, as demonstrated by GenMol (\citealp{lee2025genmoldrugdiscoverygeneralist}) and TGM-DLM (\citealp{Gong_Liu_Wu_Wang_2024}). Our polymer foundation model builds upon SELFIES-TED and hence adopts a parallel strategy for property prediction and generation.

\begin{figure*}[t] 
\centering
  \hspace*{-1cm} 
  \includegraphics[width=\textwidth]{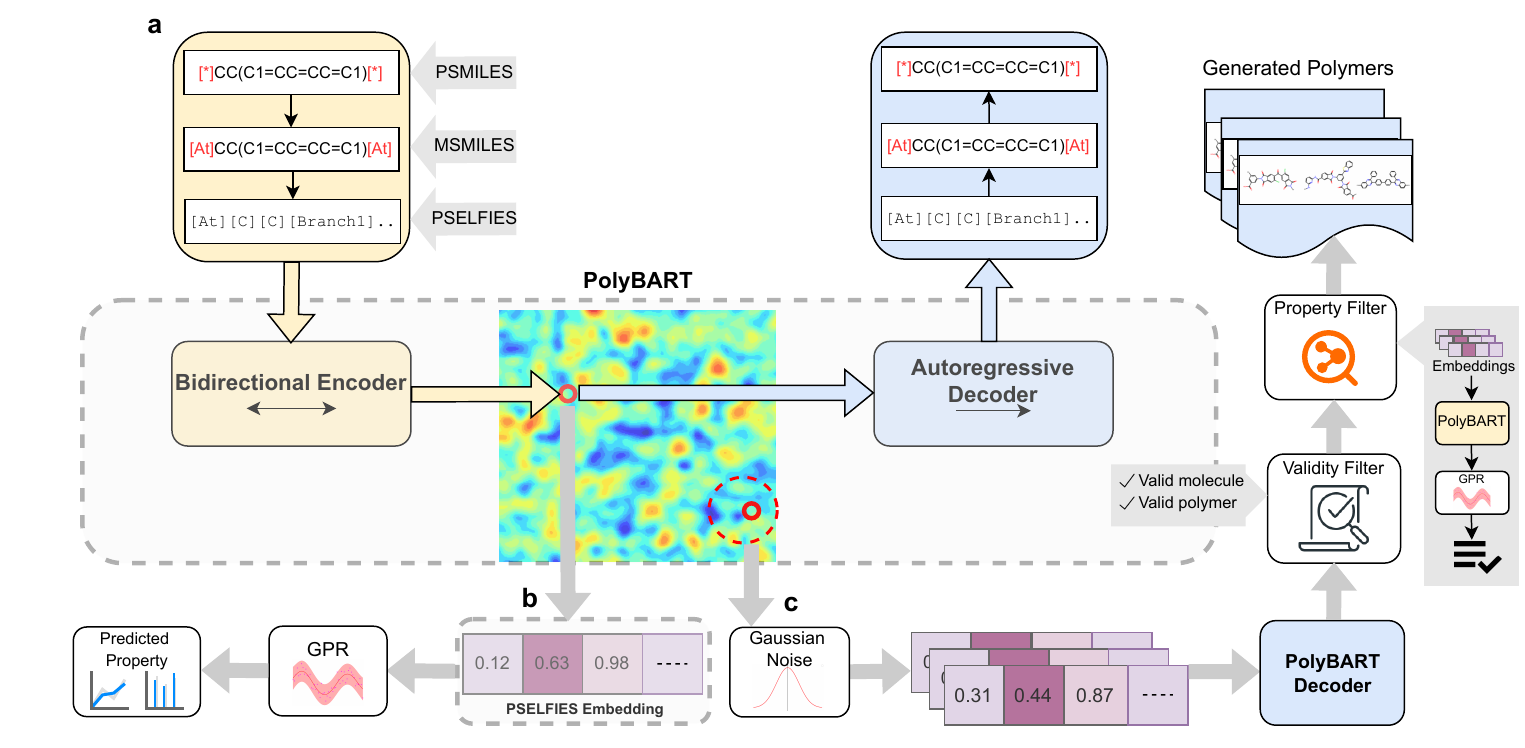}
  \caption{Overview of the polyBART pipeline. (a) The encoder-decoder model is pretrained using a Masked Language Modeling (MLM) objective to learn the language of PSELFIES and construct the latent space. (b) For property prediction, the trained encoder generates embeddings for polymers with known properties, which are then mapped to target values using Gaussian Process Regressor (GPR). (c) For generating new structures, Gaussian noise is added to the learned embeddings, and the decoder generates candidate polymers, which are subsequently filtered based on property and synthesizability criteria.}
  \label{fig:Pipeline}
\end{figure*}

\subsection{Polymer Informatics via NLP}

ML algorithms have been applied in the polymer space for quite some time (\citealp{CHEN2021100595}). Similar to molecules, a common approach in polymer informatics is to represent polymers numerically and use ML to connect these representations to points in the property space. A range of approaches, including regression algorithms (\citealp{10.1063/5.0023759}; \citealp{doi:10.1021/acs.jcim.1c01031}) and neural networks (\citealp{doi:10.1021/acs.macromol.1c00728}; \citealp{kuenneth2022bioplastic}) have been explored for this purpose. Graph neural networks (GNNs) are also widely used in polymer modeling due to their ability to capture both local and global topological features of polymer graphs. \citet{Gurnani2023Polymer} developed polyGNN, a multitask polymer GNN model for polymer property prediction, demonstrating strong performance across numerous properties. While the application of NLP in this space is relatively nascent, it has already seen significant success.  In property prediction tasks, language models are typically used to generate numerical embeddings of polymer representations, which are then mapped to properties using approaches similar to traditional polymer informatics pipelines (\citealp{doi:10.1021/acsmaterialslett.5c00054}). \citet{Kuenneth2023} developed polyBERT by pretraining deBERTA (\citealp{he2021debertadecodingenhancedbertdisentangled}) on millions of SMILES strings. Similar polymer language models have also been developed using alternative architectures, such as RoBERTa in TransPolymer (\citealp{xu2023transpolymer}) and T5 (\citealp{raffel2023exploringlimitstransferlearning}) in PolyNC (\citealp{D3SC05079C}).

For the inverse problem of generating polymer structures, early efforts have leveraged Variational Autoencoders (VAEs) (\citealp{doi:10.1021/acs.chemmater.0c03332}). However, the lack of explicit chemical syntax in VAE architectures leads to invalid generations.  Transformers can learn context-aware representations of molecular structures (\citealp{D1SC01050F}), with each attention head learning to focus on distinct substructures. By leveraging SELFIES (or PSELFIES in the present case), which guarantee validity, and language models that can capture structured sequences and syntactic patterns, we achieve state-of-the-art results in the generation of novel, valid, and property-aligned polymers.

\section{polyBART: A Polymer Foundation Model}
In this section we detail the PSELFIES representation and introduce polyBART. The overall model pipeline is shown in Figure~\ref{fig:Pipeline}.

\begin{figure*}[t] 
\centering
  \includegraphics[width=\textwidth]{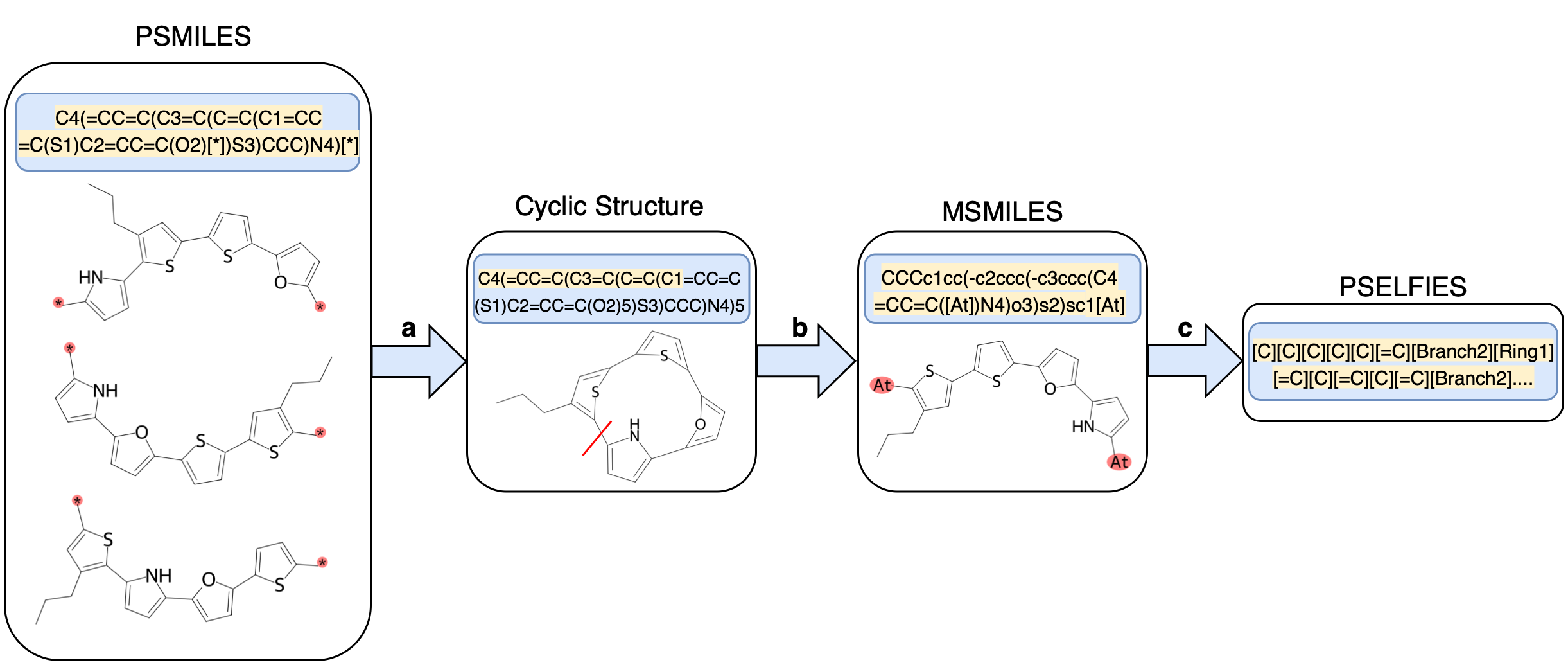}
  \caption{(a) PSMILES containing terminal [*] groups are first transformed into a cyclic structure. (b) The cyclic structure is canonicalized, and a chemically valid bond is cleaved to linearize the molecule. The resulting termini are tagged with At atoms, yielding MSMILES. (c) The MSMILES is then transformed to PSELFIES.}
  \label{fig:MSELFIES}
\end{figure*}

\subsection{PSELFIES}
\label{sec:pselfies}
 To ensure compatibility with molecular language models, we introduce PSELFIES, a novel representation of polymers. PSELFIES builds upon several representations used in cheminformatics. SMILES (Simplified Molecular Input Line Entry System) (\citealp[]{doi:10.1021/ci00057a005}) is a widely adopted notation for encoding molecules as strings. Polymer SMILES (PSMILES) extends SMILES to polymers by representing the two terminal endpoints of homopolymer repeat units with asterisks ([*]) to indicate open-chain ends. However, [*] is commonly used as a placeholder atom in molecular representations rather than specifically denoting polymer termini, and it is not yet incorporated into the SELFIES representation. 
 
 To address this limitation, we first convert PSMILES into Molecule SMILES (MSMILES), replacing [*] with actual atomic representations. The conversion begins by identifying the terminal atoms and bond types associated with [*], then joining the polymer ends to form a cyclic structure, followed by eliminating the [*]s. To ensure consistency and to remove biases arising from multiple valid PSMILES representations, we canonicalize the cyclic structure. Next, we strategically cleave a chemically valid, preferably single (non-ring) bond within the cyclic structure. Astatine (At) atoms are introduced at the newly formed termini, chosen due to their rarity in polymer chemistry, single valence character, and absence in our dataset. The resulting MSMILES representation is a linearized, chemically valid pseudo-molecular SMILES string that preserves the essential connectivity of the polymer. Finally, MSMILES is transformed into PSELFIES using the SELFIES encoder. The complete conversion pipeline is illustrated in Figure~\ref{fig:MSELFIES}. Further details on the representations described in this section are provided in Appendix~\ref{sec:representations}.

\begin{table*}[t]
  \centering
  \renewcommand{\arraystretch}{1.2} 
  \setlength{\tabcolsep}{4pt}        
  \scalebox{0.86}{%
  \begin{tabular}{lccc|cccc}
    \toprule
    \textbf{Model} & \textbf{T\textsubscript{g} (K)} & \textbf{T\textsubscript{d} (K)} & \textbf{T\textsubscript{m} (K)} & \textbf{E\textsubscript{gc} (eV)}  & \textbf{E\textsubscript{gb} (eV)}  & \textbf{E\textsubscript{ea} (eV)}  & \textbf{E\textsubscript{i} (eV)}  \\
    \midrule
    SELFIES-TED\textsubscript{large}     & 62.58 $\pm$ 1.94   &     96.98 $\pm$ 2.14                &  73.77 $\pm$ 2.97  &  0.95 $\pm$ 0.04  & 0.78 $\pm$ 0.09 & 0.37 $\pm$ 0.06 & \underline{0.50 $\pm$ 0.06}\\
    polyBERT     & \textbf{38.41 $\pm$ 1.92}   &    \textbf{67.45 $\pm$ 2.06}   &   \textbf{56.25 $\pm$ 3.09} &   \textbf{0.55 $\pm$ 0.02}  &  \underline{0.72 $\pm$ 0.06} &  0.35 $\pm$ 0.05 &  \textbf{0.49 $\pm$ 0.04} \\
    polyBART\textsubscript{small}   & \underline{39.92 $\pm$ 1.27}  &     71.56 $\pm$ 1.64                 &  \underline{57.71 $\pm$ 2.36}  &  \underline{0.60 $\pm$ 0.02}  & \textbf{0.68 $\pm$ 0.08} & 0.35 $\pm$ 0.05 & 0.51 $\pm$ 0.06 \\
    polyBART\textsubscript{large}  & 40.32 $\pm$ 1.16   &     \underline{71.44 $\pm$ 1.93}                 &  58.02 $\pm$ 2.36  &  0.61 $\pm$ 0.02  & 0.73 $\pm$ 0.07 & \textbf{0.33 $\pm$ 0.04} & \textbf{0.49 $\pm$ 0.08} \\
    \midrule
    BART\textsubscript{small}    & 43.35 $\pm$ 1.25   &  73.35 $\pm$ 1.94   &  62.03 $\pm$ 2.83 &   -  &  - &  - &  - \\
     \midrule
    LLaMA-3-8B   & 58.13  &  95.28 & 71.91  &   -  &  - &  - &  - \\
    \bottomrule
  \end{tabular}%
  }
  
  \caption{\label{citation-guide}
    Performance on the property prediction tasks. The best results are highlighted in bold and second-best are underlined.
  }
  \label{tab:prop_prediction}
\end{table*}

\subsection{Model Configuration and Details}

To develop polyBART, we construct a large-scale dataset comprising over 200 million PSELFIES strings for polymers, including 12,473 known polymers, with the remainder representing hypothetical polymers (\citealp[]{doi:10.1021/acs.jcim.3c00329}; \citealp[]{doi:10.1021/acspolymersau.3c00003}). The dataset spans a diverse range of polymer classes, including ethers, esters, amides, amines, and imides to name a few. The hypothetical polymers are generated by applying various known polymerization reactions for polyamides, polyesters, polyethers, polyurea, and polyurethane, and popular named reactions, such as Ring-Opening Metathesis Polymerization (ROMP) (\citealp{young2011chapters3and7}). We also use popular click reactions like copper-catalyzed azide-alkyne cycloaddition (CuAAC), thiol-ene/yne reactions, thiol-bromo, Diels-Alder, and SuFEx reactions, generating a diverse space of polymers while ensuring their synthetic feasibility  (\citealp{https://doi.org/10.1002/1521-3773(20010601)40:11<2004::AID-ANIE2004>3.0.CO;2-5}, \citealp{https://doi.org/10.1002/pol.20210126}).  Additional details regarding the dataset can be found in Appendix~\ref{sec:appendix}.

We choose to focus exclusively on encoder-decoder architectures, as polymer representations exhibit bidirectional dependencies, meaning each token's presence is influenced by both its preceding and succeeding context. This necessitates bidirectional attention and the use of Masked Language Modeling (MLM) rather than Causal Language Modeling (CLM), ruling out decoder-only models in our case. Given its architectural compatibility and vocabulary aligned with our input format, SELFIES-TED is the most suitable molecular foundation model for transfer learning to our task.

The SELFIES-TED model, based on the BART architecture, comes in two variants, SELFIES-TED\textsubscript{small} with 2.2 million parameters (2 encoder-decoder layers, 4 attention heads) and SELFIES-TED\textsubscript{large} with 358 million parameters (12 encoder-decoder layers, 16 attention heads), both trained on SELFIES molecular representations. We extend both variants of SELFIES-TED to the polymer space via continued pretraining on our polymer dataset, with polyBART\textsubscript{small} trained on \textasciitilde200 million polymers and polyBART\textsubscript{large} trained on a subset of 25 million samples. Prior to training, we found that the SELFIES-TED\textsubscript{small} tokenizer did not include \texttt{[At]} in its vocabulary, which we add with a randomly initialized embedding. 

We adopt the same training strategy as SELFIES-TED, using a denoising objective with 15\% of the tokens in the input sequence randomly masked. Detailed descriptions of the model architecture and training setup can be found in Appendix~\ref{sec:architecture}. Ultimately, our approach allows us to leverage the extensive chemical knowledge encoded in the base models, pretrained on billions of molecules, while progressively specializing the models to learn the representations of polymers. Through further training on millions of PSELFIES strings, polyBART thus obtains a deep understanding of the grammar and syntax that govern the polymer chemical language.

\section{Computational Experiments}
In this section, we present results demonstrating the effectiveness of polyBART on: (i) property prediction, (ii) hypothetical polymer generation, and (iii) property-guided never-seen-before polymer generation. We also highlight the ability of PSELFIES in leveraging pretrained knowledge through a comparative study.

\subsection{Property Prediction}

\textbf{Task}: Polymer property prediction is formulated as a supervised learning task in our pipeline and accomplished as follows: first, all the PSMILES strings from the property datasets are converted to their PSELFIES representations. These are then passed through the polyBART encoder to generate numerical embeddings. To reduce dimensionality, we apply Principal Component Analysis (PCA) to the resulting PSELFIES embeddings. The reduced representations are then used to train a supervised learning model, specifically, Gaussian Process Regression (GPR) with a Radial Basis Function (RBF) kernel. We use a 5-fold Cross-Validation (CV) strategy to assess the performance and generalizability of our GPR models.

\textbf{Datasets}: We evaluate polyBART on critical thermal properties of polymers, namely, the glass transition temperature (T\textsubscript{g}), thermal degradation temperature (T\textsubscript{d}), and melting temperature (T\textsubscript{m}) (\citealp{Kuenneth2021Polymer}). Thermal property data is collected from the PolyInfo repository (\citealp{6076416}) and is derived from experimental measurements reported in the existing literature. We also assess performance on electronic properties, including linear chain bandgap (E\textsubscript{gc}), bulk polymer bandgap (E\textsubscript{gb}), electron affinity (E\textsubscript{ea}), and ionization energy (E\textsubscript{i}) values computed using Density Functional Theory (DFT). Additional information on the datasets is detailed in Appendix~\ref{sec:properties}.

\textbf{Baselines}: We compare the performance of polyBART to the baseline SELFIES-TED model. Due to the absence of the \texttt{[At]} token in SELFIES-TED\textsubscript{small}, our comparison is restricted to SELFIES-TED\textsubscript{large}. In addition, we benchmark polyBART with polyBERT, current state-of-the-art transformer-based model for polymer property prediction. It is important to note that, given the novel bidirectional capabilities of polyBART, we are limited in comparison to unidirectional models developed for property prediction. To ensure a fair comparison of embedding quality, we use GPR across all models. We additionally compare polyBART with LLaMA, with details regarding the fine-tuning experiment given in Appendix~\ref{sec:llama}.

\textbf{Results}: Table~\ref{tab:prop_prediction} presents the performance of polyBART\textsubscript{small} and polyBART\textsubscript{large}, reporting the average RMSE across five CV test sets along with the corresponding standard deviations. We observe that our bidirectional polyBART performs on par with, and in many cases surpasses, polyBERT. Notably, while matching the predictive accuracy of the unidirectional polyBERT, the bidirectional polyBART offers the added advantage of generating entirely new polymer candidates not present in the training data. Additionally, polyBART\textsubscript{small} and polyBART\textsubscript{large} outperform the baseline SELFIES-TED\textsubscript{large}, achieving average RMSE reductions of 19.4\% and 19.7\%, respectively. These consistent performance gains underscore the ability of PSELFIES to specialize molecular language models to the polymer domain. We also notice that polyBART outperforms LLaMA-3 on thermal property prediction. Overall, our results confirm the capability of polyBART as a powerful polymer property predictor.

\begin{figure*}[t] 
\centering
 \hspace*{-1cm} 
  \includegraphics[width=\textwidth]{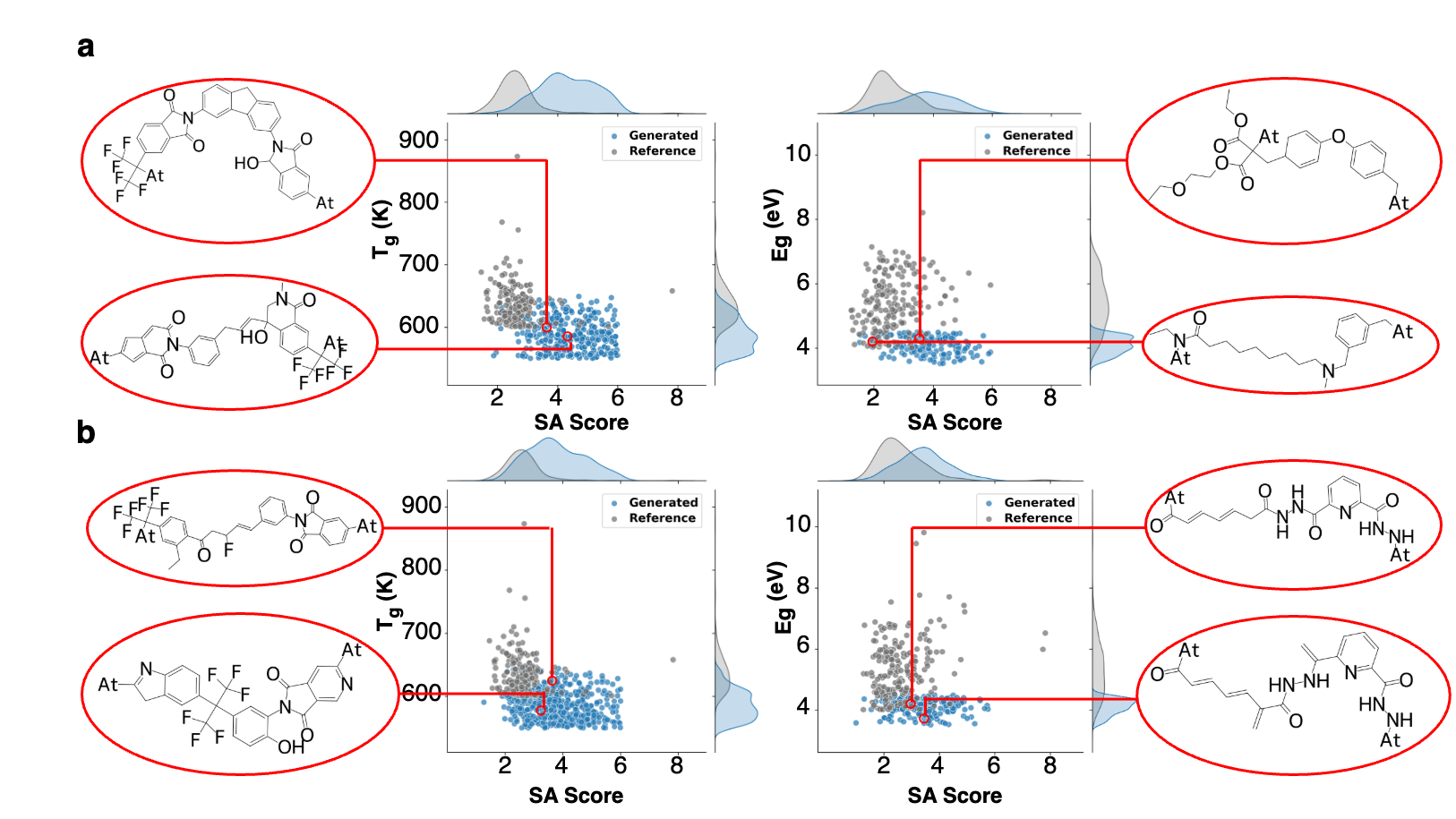}
  \caption{Representative chemical structures generated by (a) polyBART\textsubscript{small} and (b) polyBART\textsubscript{large}, selected for their high T\textsubscript{g}, high E\textsubscript{g}, and ease of synthesizability. The selected examples highlight polyBART's ability to generate thermally and electronically robust polymers.}
  \label{fig:Structures}
\end{figure*}

\subsection{Impact of Transfer Learning}

To further demonstrate the importance of PSELFIES in adapting molecular foundation models to the polymer space, we perform a comparative study. Specifically, we develop a model, BART\textsubscript{small}, which shares the same architecture, training data, and configuration as polyBART\textsubscript{small}, but is trained from scratch rather than initialized from SELFIES-TED. This setup allows us to isolate the effect of transfer learning using PSELFIES. 

We compare the performance of BART\textsubscript{small} against polyBART\textsubscript{small} on property prediction tasks across the three thermal properties, with results summarized in Table~\ref{tab:prop_prediction}. We observe that polyBART\textsubscript{small} consistently outperforms BART\textsubscript{small} in all tasks. These performance gains highlight the critical role of our PSELFIES framework. By leveraging pretrained knowledge from SELFIES-TED, polyBART benefits from a strong initialization, leading to significant improvements over training from scratch. Overall, we confirm that rather than starting from random weights, initializing with a pretrained molecular model via PSELFIES provides an effective foundation for developing polymer language models.

\begin{table*}[t]
  \centering
  \scalebox{0.95}{%
  \begin{tabular}{lccccccc}
    \toprule
    \textbf{Model} & \textbf{Molecule Validity} & \textbf{Polymer Validity} & \textbf{Novelty } & \textbf{FCD}  & \textbf{IntDiv\textsubscript{1} } & \textbf{IntDiv\textsubscript{2} } \\
    \midrule
    polyBART\textsubscript{small}      & 1.000   &     0.913                 &  0.867  &  3.968    &  0.878  &  0.875 \\
    polyBART\textsubscript{large}     & 1.000   &     0.985                 &  0.793 &  1.833    &  0.872   &  0.868  \\
    \bottomrule
  \end{tabular}%
  }
  \caption{\label{citation-guide}
    Generative performance of polyBART\textsubscript{small} and polyBART\textsubscript{large} across key evaluation metrics.
  }
  \label{tab:generation}
\end{table*}

\begin{table*}
  \centering
  \begin{tabular}{ccc|cccccc}
    \toprule
    \multirow{2}{*}{\textbf{Metric}}  &
    \multicolumn{2}{c}{\textbf{T\textsubscript{g}}} &
    \multicolumn{2}{c}{\textbf{E\textsubscript{gc}}}  \\
    & \textbf{polyBART\textsubscript{small}} & \textbf{polyBART\textsubscript{large}} & \textbf{polyBART\textsubscript{small}} & \textbf{polyBART\textsubscript{large}} \\
    \midrule
    \# Valid Generations            & 743 & 1131 & 739 & 1514\\
    \# Novel            & 715 & 933 & 558 & 660 \\
    \# Property Filtered & 396   & 543   & 168   & 190 \\
    \# SA Score Filtered     & \textbf{353} & \textbf{511} & \textbf{160} & \textbf{189} \\
    \bottomrule
  \end{tabular}
  \caption{\label{property-model-count-horizontal}
    Performance of polyBART\textsubscript{small} and polyBART\textsubscript{large} on T\textsubscript{g} and E\textsubscript{gc} guided generation of novel polymer candidates, based on 183 and 200 reference polymers, respectively.
  }
  \label{tab:generation2}
\end{table*}

\subsection{Hypothetical Polymer Generation}

polyBART's generative capability, enabled by its decoder, is a key innovation of our work, setting it apart from existing polymer language models (\citealp[]{doi:10.1021/acsmaterialslett.5c00054}), which have been limited to unidirectional property prediction. To generate novel polymer structures, the PSELFIES sequence is first passed through the encoder to obtain a latent embedding. To explore the latent space, we apply Gaussian noise to this embedding using an n-fold sampling strategy: noise is added n times to generate n noised embeddings. These noisy embeddings are then decoded one by one using the polyBART decoder, and if a valid polymer is generated within these samples, it is retained. A grid search is used to tune the noise level for novelty and validity, details of which can be found in Appendix~\ref{sec:noise}.

We evaluate the generative capabilities of polyBART using a test set of 10,000 PSELFIES, all of which are unseen during the pretraining phase. We begin by examining the validity of the generated polymer structures. Specifically, we introduce two measures of validity: (i) molecule validity and (ii) polymer validity. Molecule validity checks whether the generated PSELFIES correspond to a chemically valid molecule, and is determined using RDKit. Polymer validity assesses whether the generated PSELFIES represent a structurally valid polymer by examining the placement and single-bond character of the \texttt{[At]} tokens. Since the [At] tokens denote the endpoints of the polymer repeat unit, a valid polymer must contain exactly two such tokens, each with a valency of one. The polymer validity metric enforces this requirement. Next, we assess the novelty of the valid generated structures, where a structure is considered novel if it passes both validity tests and is different from all input examples. Finally, we assess the quality of the generated structures using Fréchet ChemNet Distance (FCD) (\citealp{preuer2018frechetchemnetdistancemetric}) and Internal Diversity (IntDiv\textsubscript{p}) metrics (\citealp{benhenda2017chemganchallengedrugdiscovery}). FCD quantifies the similarity
between the distributions of the generated and input structures based on the Euclidian distance between their respective embeddings. A higher FCD score indicates a greater divergence between the two distributions, suggesting that the generated structures differ from the input set. IntDiv\textsubscript{p} evaluates the internal diversity of the generated structures, capturing the model's tendency to produce structurally distinct outputs. We report both IntDiv\textsubscript{1} (p = 1) and IntDiv\textsubscript{2} (p = 2) scores, where higher values of these scores reflect greater diversity in the generated structures. Although these metrics were originally designed for molecules, by representing polymers as pseudo-molecules using PSELFIES, we can extend them to our case. See Appendix~\ref{sec:metrics} for a detailed description of the metrics.

Results for the generation task are provided in Table~\ref{tab:generation}. We find that polyBART\textsubscript{small} and polyBART\textsubscript{large} generate valid molecules in 100\% of cases, and valid polymers in 91\% and 98\% of cases, respectively, demonstrating that the models have effectively learned the underlying grammar and structural rules in writing polymers. High novelty scores demonstrate polyBART’s ability to explore the latent space and generate thousands of never-seen-before structures. The FCD and IntDiv\textsubscript{p} scores of the generated distributions confirm the models' capabilities to generate a broad and diverse range of polymer structures.

\subsection{Property-conditioned Polymer Generation}

Beyond general structure generation, a key capability of polyBART is its ability to generate novel polymers conditioned on desired target properties. At a high level, our approach enables property-guided generation by mapping the latent space of polymer embeddings to the property space and exploring regions with desired property values. We begin by identifying polymers that fall within a specified target range for a given property. These polymers are then passed through polyBART's encoder to obtain embeddings. As in the general generation setting, we apply an n-fold Gaussian noise sampling strategy: each embedding is perturbed n times to generate n candidate embeddings, which are subsequently decoded and checked for validity. In this way, we are able to effectively explore neighborhoods in the latent space with desirable property values. The resulting candidate polymers are subjected to two filtering steps: (i) property filtering and (ii) Synthetic Accessibility (SA) score (\citealp{ertl2009sascore}) filtering. To filter based on property, the generated candidates are passed through the GPR models developed for property prediction, and only those within a threshold range of the target property are retained. To assess synthesizability, we compute the SA score, a measure ranging from 1 (easy to make) to 10 (very difficult to make). For this, PSELFIES strings are converted to MSMILES, with the At atom replaced by hydrogen (H). The SA score is then calculated, and candidates with a score of 6 or lower are retained as viable candidates. Additional details are provided in Appendix~\ref{sec:sa_score}. 

We use polyBART to generate novel polymer structures under two extreme conditions: (i) high temperatures and (ii) high electric fields. We make the assumption that these behaviors correlate with two critical polymer properties, specifically,  T\textsubscript{g} and E\textsubscript{gc}. Accordingly, we define our target property thresholds as (i) T\textsubscript{g} > 600 K and (ii) E\textsubscript{gc} > 4 eV. High T\textsubscript{g} polymers are sought for their thermal integrity at elevated temperatures, while polymers with high E\textsubscript{gc} values provide broad electrical stability with minimal dielectric loss, making them promising candidates for high-energy density capacitor dielectric applications (\citealp{article}; \citealp{https://doi.org/10.1002/adma.201805672}; \citealp{gurnani2024ai}).

As shown in Table~\ref{tab:generation2}, polyBART\textsubscript{small} and polyBART\textsubscript{large} are able to successfully generate hundreds of novel polymers that not only meet the desired property criteria but are also predicted to be synthetically accessible. Examples of novel structures generated by our models are illustrated in Figure~\ref{fig:Structures}. Additionally, visualizations of the property-conditioned latent space can be found in Appendix~\ref{sec:latent_space}.

\begin{figure}[H]
\centering
    \centering
    \includegraphics[width=0.75\linewidth]{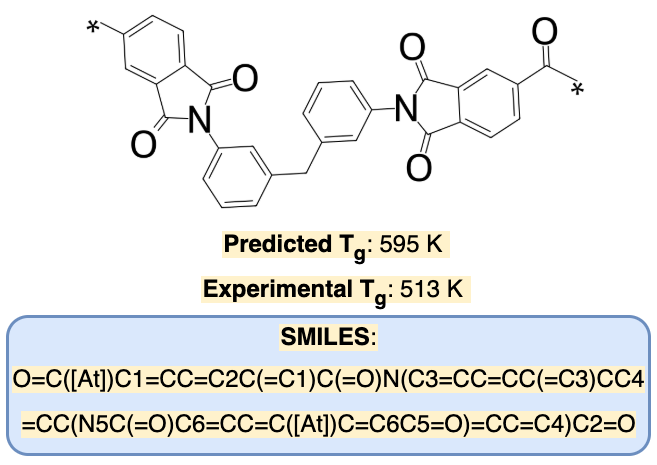}
    \caption{A high T\textsubscript{g} polymer structure generated by polyBART and synthesized in the lab.}
    \label{fig:struct}
\end{figure}

\section{Validation of Design via Laboratory Synthesis and Testing}

Given the real-world implications of our work, it is important to extend our evaluation of polyBART beyond computational benchmarks. Thus, we perform experimental synthesis and characterization of a polymer candidate predicted by polyBART\textsubscript{small}. We select one polymer with a high predicted T\textsubscript{g} and a favorable SA score and successfully synthesize it in the lab. Following synthesis, we perform T\textsubscript{g} measurements to evaluate its thermal properties. The experimental protocol is detailed in Appendix~\ref{sec:synthesis}. Figure~\ref{fig:struct} presents the chemical structure of the synthesized polymer, along with the predicted and experimentally measured T\textsubscript{g} values. We confirm that the experimentally observed T\textsubscript{g} aligns well with the prediction. This successful synthesis of a high-T\textsubscript{g} polymer firmly establishes polyBART's ability to guide real-world polymer design. 

\section{Conclusion}

In this work, we propose PSELFIES, a novel representation of polymers that enables the transfer of learned knowledge from molecular language models to the polymer space. We develop polyBART\textsubscript{small} and polyBART\textsubscript{large}, two polymer foundation models created through the continued pretraining of SELFIES-TED\textsubscript{small} and SELFIES-TED\textsubscript{large}, respectively. Our results show that polyBART performs comparably to state-of-the-art approaches in property prediction. Moreover, polyBART uniquely combines property prediction with generative design in a single unified framework and consistently produces numerous novel polymer structures. Generated candidates not only satisfy target property criteria but also exhibit synthetic accessibility. By successfully synthesizing one such predicted polymer and experimentally confirming our predictions, we illustrate that language models can reliably drive real-world advances in polymer design. Our work enables true bidirectional translation between polymer structures and properties and highlights the potential of language models in advancing polymer informatics.

\section{Limitations}
While PSELFIES successfully enables the adaptation of the SELFIES-TED molecular language model to the polymer space, its generalizability to other molecular foundation models, based on different architectures and pretraining strategies remains yet to be seen. Future work is needed to evaluate how well PSELFIES integrates with alternative models and whether similar performance gains can be achieved across methods.

Additionally, while polyBART demonstrates strong performance on thermal and electronic properties, its applicability to a broader range of polymer properties and simultaneous optimization of multiple properties must be explored to fully determine its predictive and generative power. Finally, polyBART is currently limited to homopolymers and does not yet support the representation or generation of copolymers, polymer blends, or polymer composites/formulations. These aspects will be the subject of future inquiry.

\section{Acknowledgements}
The authors thank the Office of Naval Research (ONR) for financial sponsorship of this work through Grant N00014-23-1-2279. The authors are also grateful for the usage of the Gutekunst lab for synthesis and characterization of one of the designed polymers.

\bibliography{custom}

\clearpage
\appendix
\section*{Appendix}
\setcounter{figure}{4}  
\setcounter{table}{3}  
\setcounter{page}{12}
\section{Representations}
\label{sec:representations}
\begin{table}[H]
\centering
\captionsetup{width=\columnwidth}  
\begin{tabular}{l p{0.47\columnwidth}}
\toprule
\textbf{Representation} & \textbf{Description} \\
\midrule
SMILES              &  Simplified Molecular Input Line Entry System is a popular notation for encoding molecules as strings.\\
PSMILES             & Polymer SMILES is an extension of SMILES for polymers that uses asterisks ([*]) to indicate endpoints of homopolymer repeat units. \\
MSMILES             & Molecule SMILES is a pseudo-molecular SMILES derived from PSMILES by ring closure, canonicalization, and bond cleavage. \\
SELFIES             & Self-Referencing Embedded Strings is a robust representation of molecules that guarantees the validity of every generation. \\
PSELFIES            & Polymer SELFIES is a novel polymer-specific extension of SELFIES that is introduced in this work and used to develop polyBART.\\
\bottomrule
\end{tabular}
\caption{Overview of molecular and polymer representations referenced in this work.}
\label{tab:representations}
\end{table}

\section{Pre-training Dataset}
\label{sec:appendix}
We characterize the chemical diversity of our dataset of approximately 210 million polymers (total count = 210,645,750) that we use for pre-training polyBART. Polymers are categorized based on the presence of 40 functional groups. Classification is performed by analyzing the entire polymer structure, allowing a single polymer to be assigned to multiple categories.

\begin{table}[H]
\centering
\renewcommand{\arraystretch}{1.2}
\begin{tabular}{l r}
\toprule
\textbf{Reaction Class} & \textbf{Percentage} \\
\midrule
Ether              &  97.2 \\
Amide    & 72.5 \\
Amine    &  67.1\\
Acetal   &  34.5\\
Halogenated  &  33.8\\
Ester    &  23.7\\
Thioether    &  23.6\\
Allyl    &  21.2\\
Hydroxyl     &  17.7\\
Sulfoxide    &  16\\
Urea     &  11.4\\
Imide    &  8.3\\
Nitrile  &  6\\
Carboxylic acid  & 5.2 \\
Urethane     &  4.9\\

\bottomrule
\end{tabular}
\label{tab:reaction-classes}
\caption{Top 15 polymer classes in the pretraining data.}
\end{table}

\section{Architecture and Training Setup}
\label{sec:architecture}
\begin{table}[H]
\centering
\renewcommand{\arraystretch}{1.2}
\resizebox{\columnwidth}{!}{
\begin{tabular}{lcc}
\toprule
\textbf{Configuration Parameter} & \textbf{polyBART\textsubscript{small}} & \textbf{polyBART\textsubscript{large}} \\
\hline
d\textsubscript{model} & 256 & 1024 \\
Encoder Layers & 2 & 12 \\
Decoder Layers & 2 & 12 \\
Encoder FFN Dim. & 256 & 4096 \\
Decoder FFN Dim. & 256 & 4096 \\
Encoder Attention Heads & 4 & 16 \\
Decoder Attention Heads & 4 & 16 \\
\bottomrule
\end{tabular}
}
\caption{Model architecture configurations for polyBART\textsubscript{small} and polyBART\textsubscript{large} as following SELFIES-TED\textsubscript{small} and SELFIES-TED\textsubscript{large}.}
\end{table}

The loss function for training is given by:
\[
\mathcal{L}_{\text{denoise}} = - \sum_{t=1}^{T} \log P(y_t \mid y_{<t}, \tilde{x}; \theta)
\]
where $y_t$ is the $t$-th token in the output sequence $y$, $y_{<t}$ is the sequence of tokens before position $t$, $\tilde{x}$ is the corrupted version of the original input, $x$, $\theta$ represents the model parameters, and $P(y_t \mid y_{<t}, \tilde{x}; \theta)$ is the model's predicted probability of the token $y_t$ conditioned on the previously generated tokens and the corrupted input. 

polyBART\textsubscript{small} is trained on \textasciitilde200 million PSELFIES for 6 epochs using a per-device batch size of 128. Due to computational limitations, polyBART\textsubscript{large} is trained on 25 million samples for 5 epochs with a per-device batch size of 64. All training runs are conducted on 2 NVIDIA L40S GPUs.

\section{Property Datasets}
\label{sec:properties}
We evaluate polyBART on T\textsubscript{g}, T\textsubscript{d}, T\textsubscript{m}, E\textsubscript{gc}, E\textsubscript{gb}, E\textsubscript{ea}, and E\textsubscript{i}. Histograms of the value distributions for each property are shown below. Thermal properties are obtained from PolyInfo at \url{https://polymer.nims.go.jp/}. Electronic properties are accessible through the polyVERSE repository (\url{https://github.com/Ramprasad-Group/polyVERSE/tree/main}) and the Khazana polymer database (\url{https://khazana.gatech.edu/dataset/}).

\subsection{Thermal Properties}
\begin{figure}[H]
\centering
\begin{subfigure}[b]{0.9\columnwidth}
    \centering
    \includegraphics[width=\linewidth]{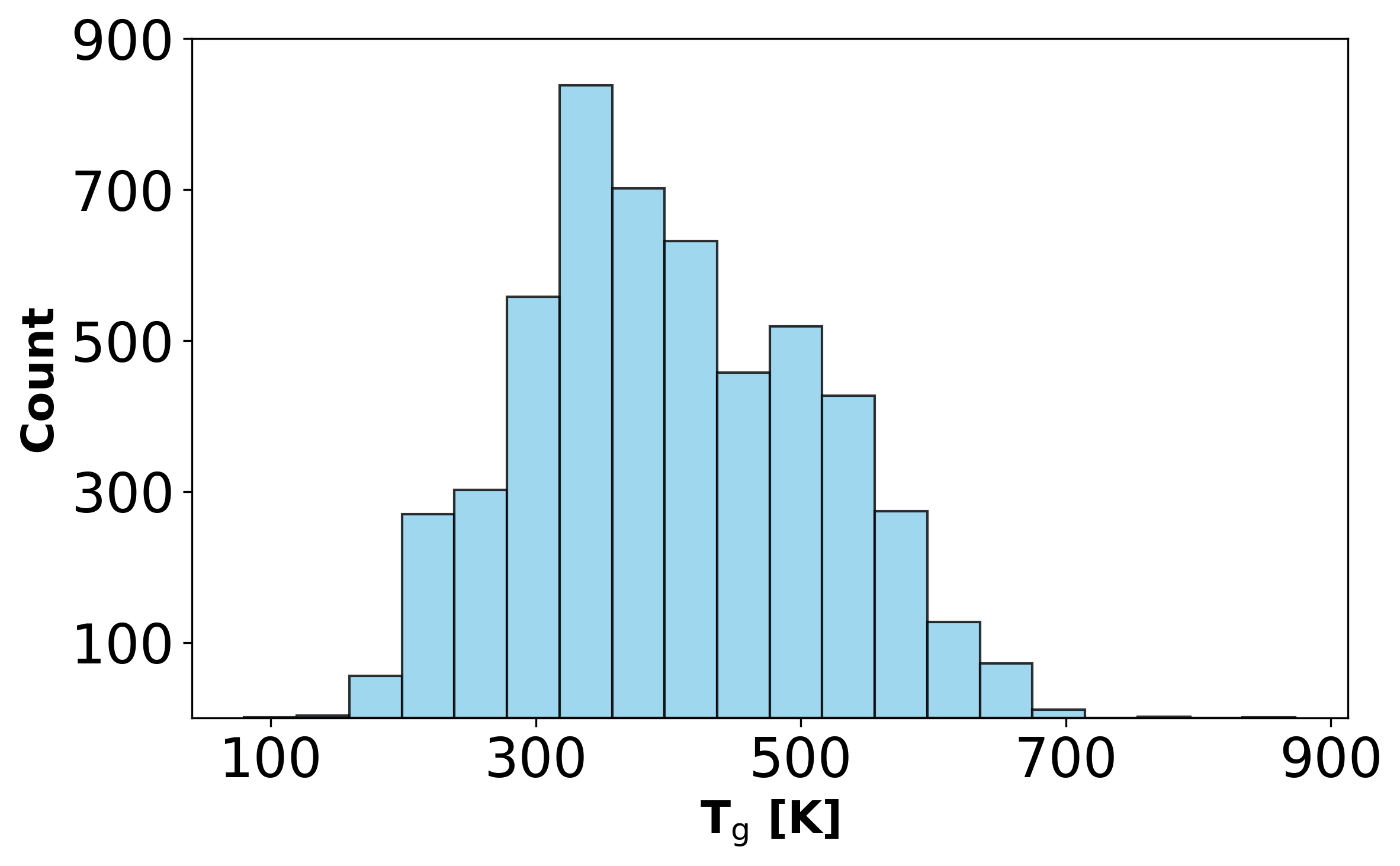}
    \label{fig:latent-tg}
\end{subfigure}

\vspace{1em}  

\begin{subfigure}[b]{0.9\columnwidth}
    \centering
    \includegraphics[width=0.96\linewidth]{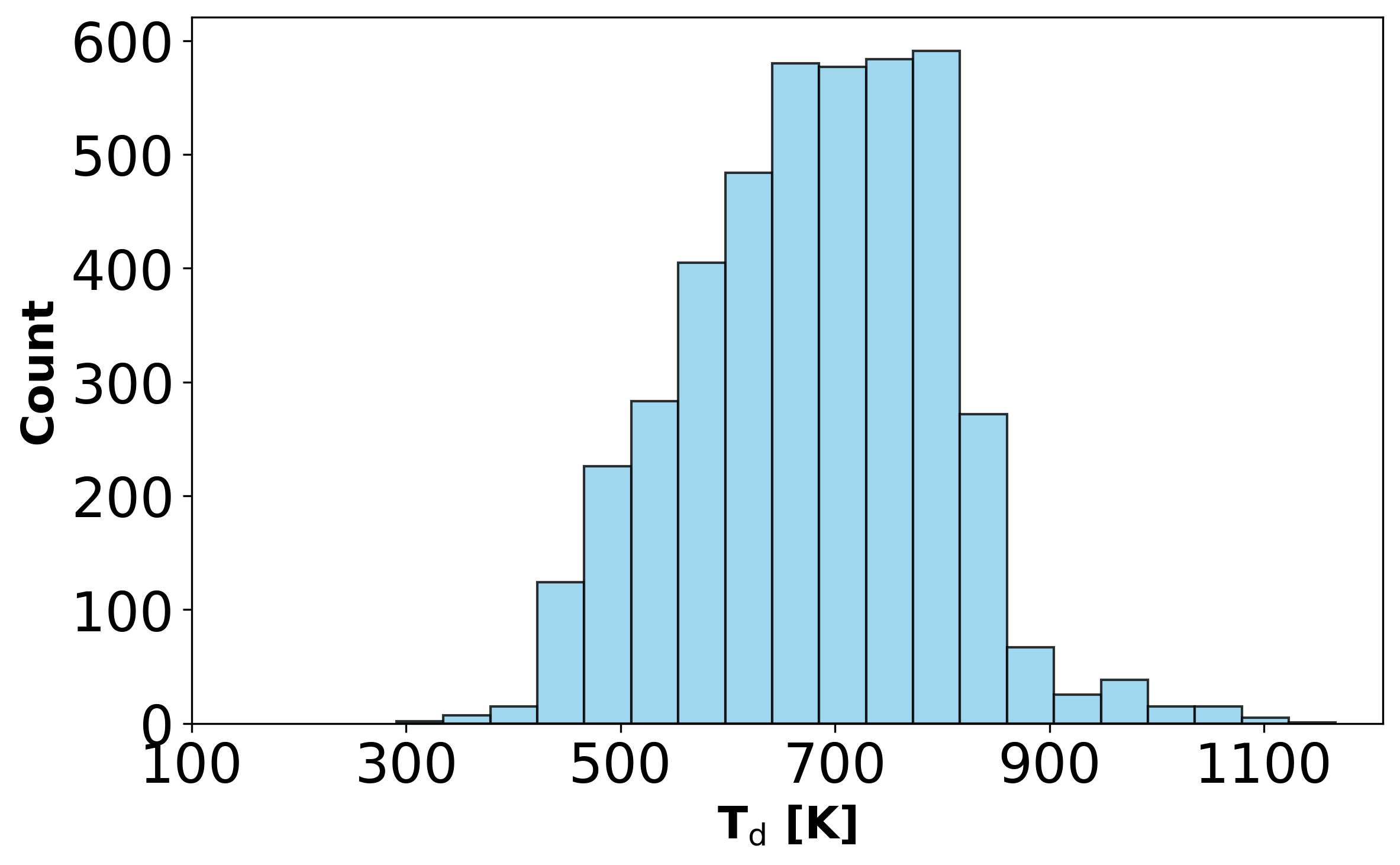}
    \label{fig:latent-eg}
\end{subfigure}

\vspace{1em} 
\begin{subfigure}[b]{0.9\columnwidth}
    \centering
    \includegraphics[width=\linewidth]{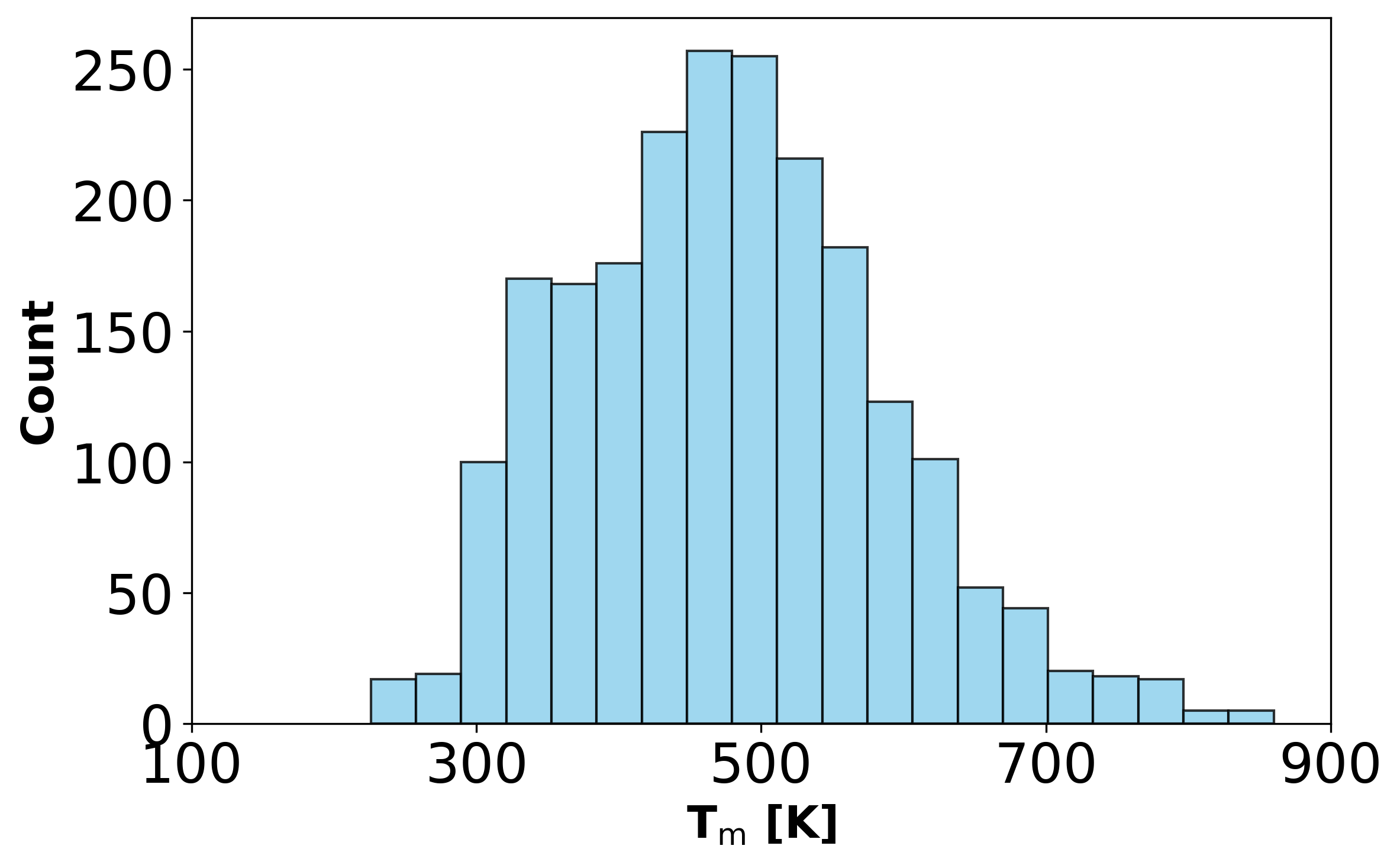}
    \label{fig:latent-eg}
\end{subfigure}

\caption{Distribution of thermal properties.}
\label{fig:latent-space-merged}
\end{figure}

\subsection{Electronic Properties}
\begin{figure}[H]
\centering
\begin{subfigure}[b]{0.9\columnwidth}
    \centering
    \includegraphics[width=\linewidth]{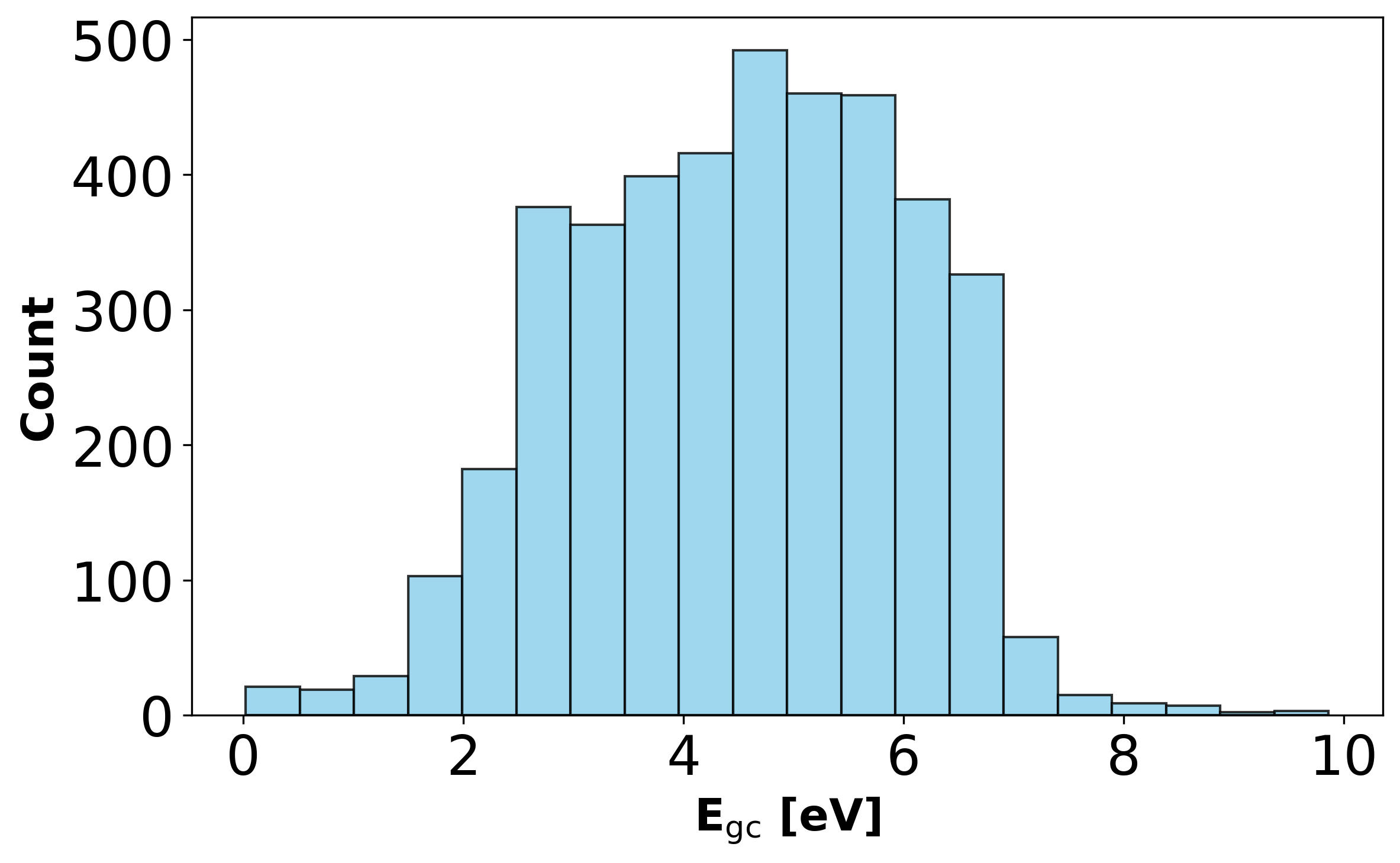}
    \label{fig:latent-tg}
\end{subfigure}

\begin{subfigure}[b]{0.9\columnwidth}
    \centering
    \includegraphics[width=\linewidth]{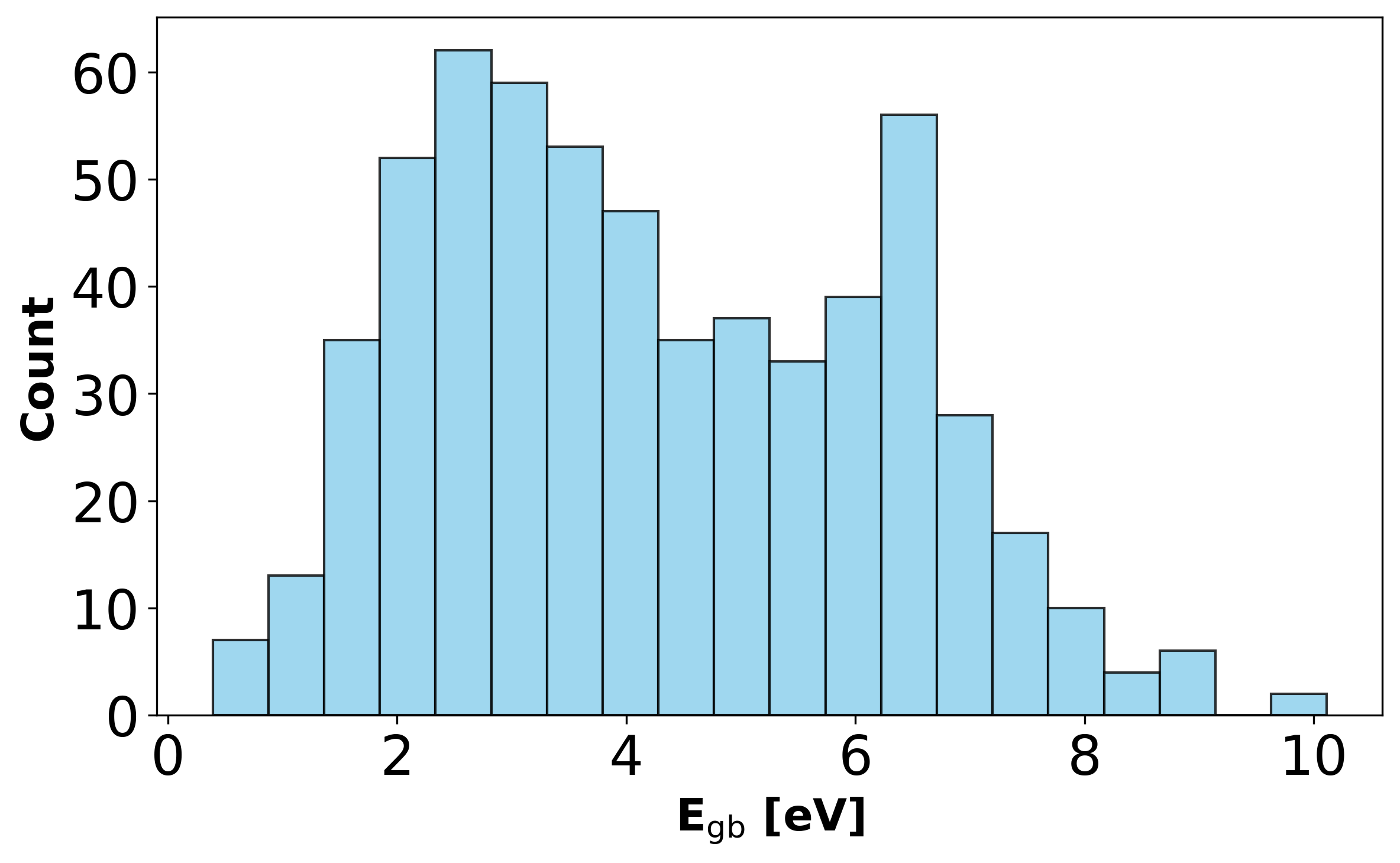}
    \label{fig:latent-eg}
\end{subfigure}

\begin{subfigure}[b]{0.9\columnwidth}
    \centering
    \includegraphics[width=\linewidth]{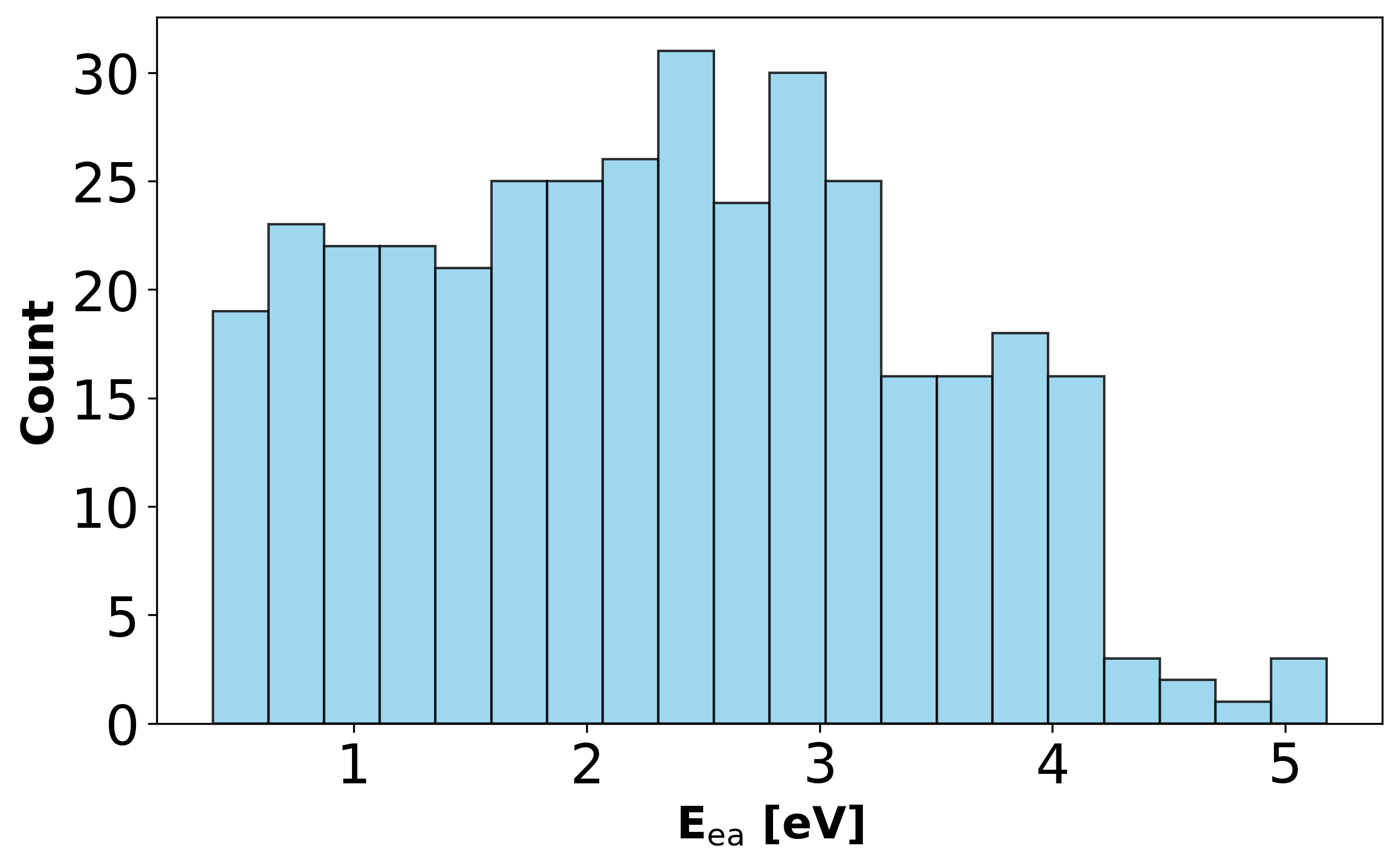}
    \label{fig:latent-eg}
\end{subfigure}

\begin{subfigure}[b]{0.9\columnwidth}
    \centering
    \includegraphics[width=\linewidth]{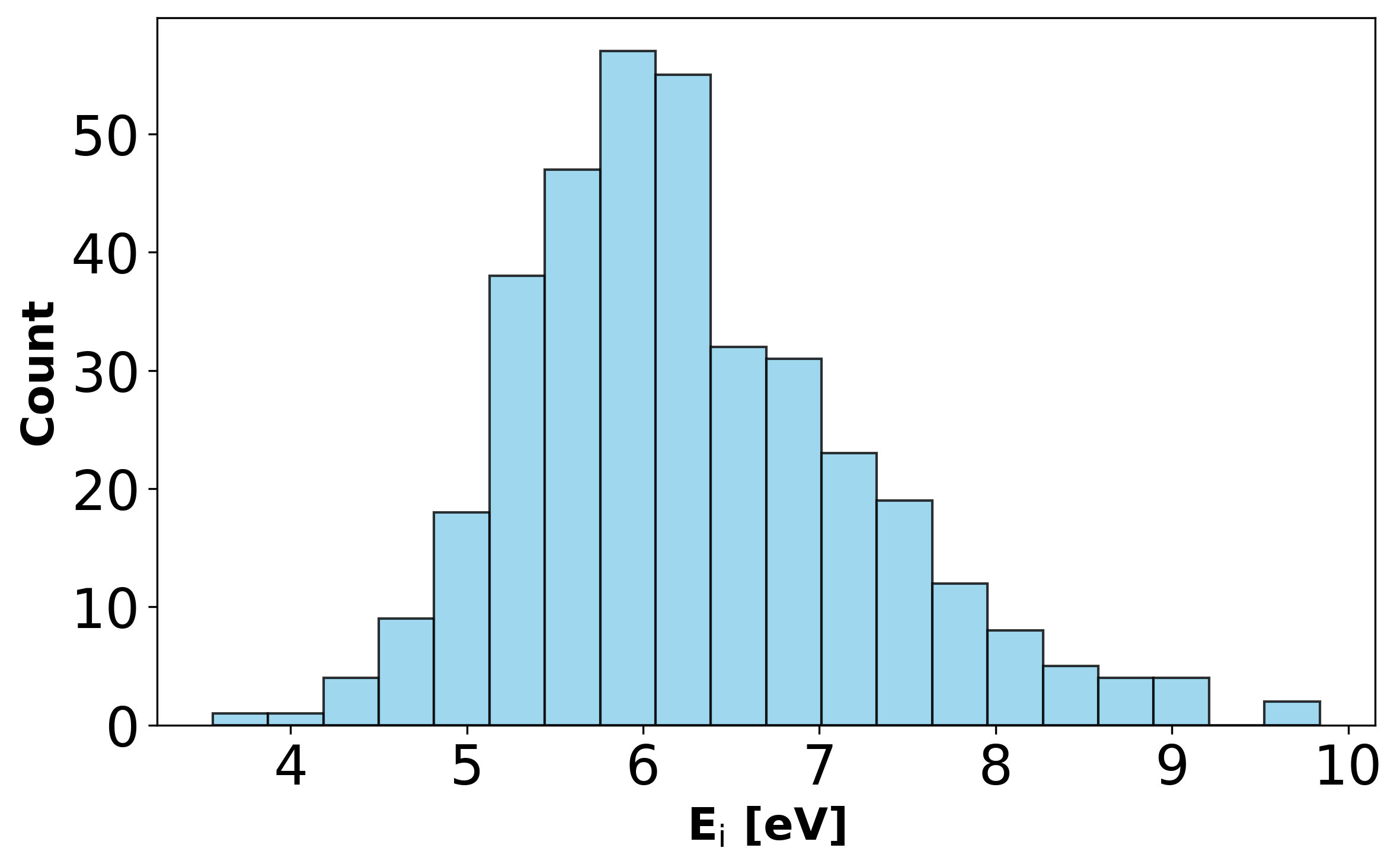}
    \label{fig:latent-eg}
\end{subfigure}

\caption{Distribution of electronic properties.}
\label{fig:latent-space-merged}
\end{figure}

\section{Comparison with LLaMA}
\label{sec:llama}

To enable a comparison between polyBART and an LLM, we evaluate both approaches on the task of thermal property prediction, using the 8B-parameter LLaMA-3 model as the representative LLM baseline. Because the available electronic-property datasets are comparatively small, we limit this comparison to thermal properties.

We fine-tune LLaMA to follow a simple question–answer protocol that maps SELFIES to scalar property values. Specifically, each training instance is formatted as:

\textit{Prompt}: What is the \{property name\} of the polymer with SELFIES: \{selfies\}?\\
\textit{Response}: \{property value\}

We fine-tune LLaMA with LoRA ($r=8$, $\alpha=8$) and early stopping. At inference time, we employ deterministic decoding (temperature $=0$, no sampling) to produce a single numeric answer.

\section{Optimal Noise Level}
\label{sec:noise}
To identify the optimal level of noise for property-conditioned generation, we analyze the effect of varying noise levels on two metrics: validity and novelty. The following plots summarize these metrics for polyBART across different noise levels. Based on our observations we choose a noise level of 0.75 for polyBART\textsubscript{small} and 1.75 for polyBART\textsubscript{large}.

\begin{figure}[H]
\centering
\begin{subfigure}[b]{0.95\columnwidth}
    \centering
    \includegraphics[width=\linewidth]{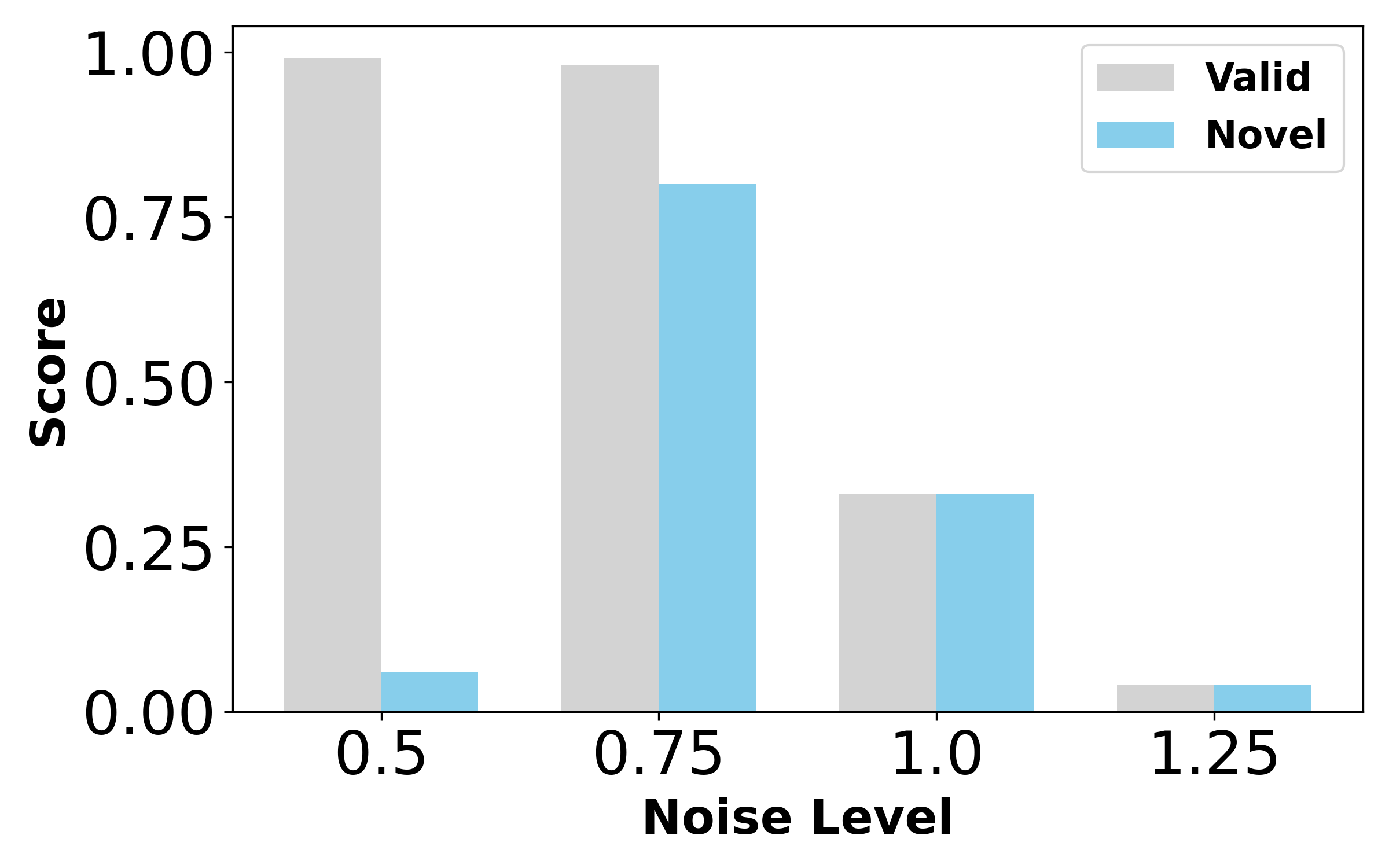}
    \caption{polyBART\textsubscript{small}}
    \label{fig:latent-tg}
\end{subfigure}

\vspace{0.5em}  
\begin{subfigure}[b]{0.95\columnwidth}
    \centering
    \includegraphics[width=\linewidth]{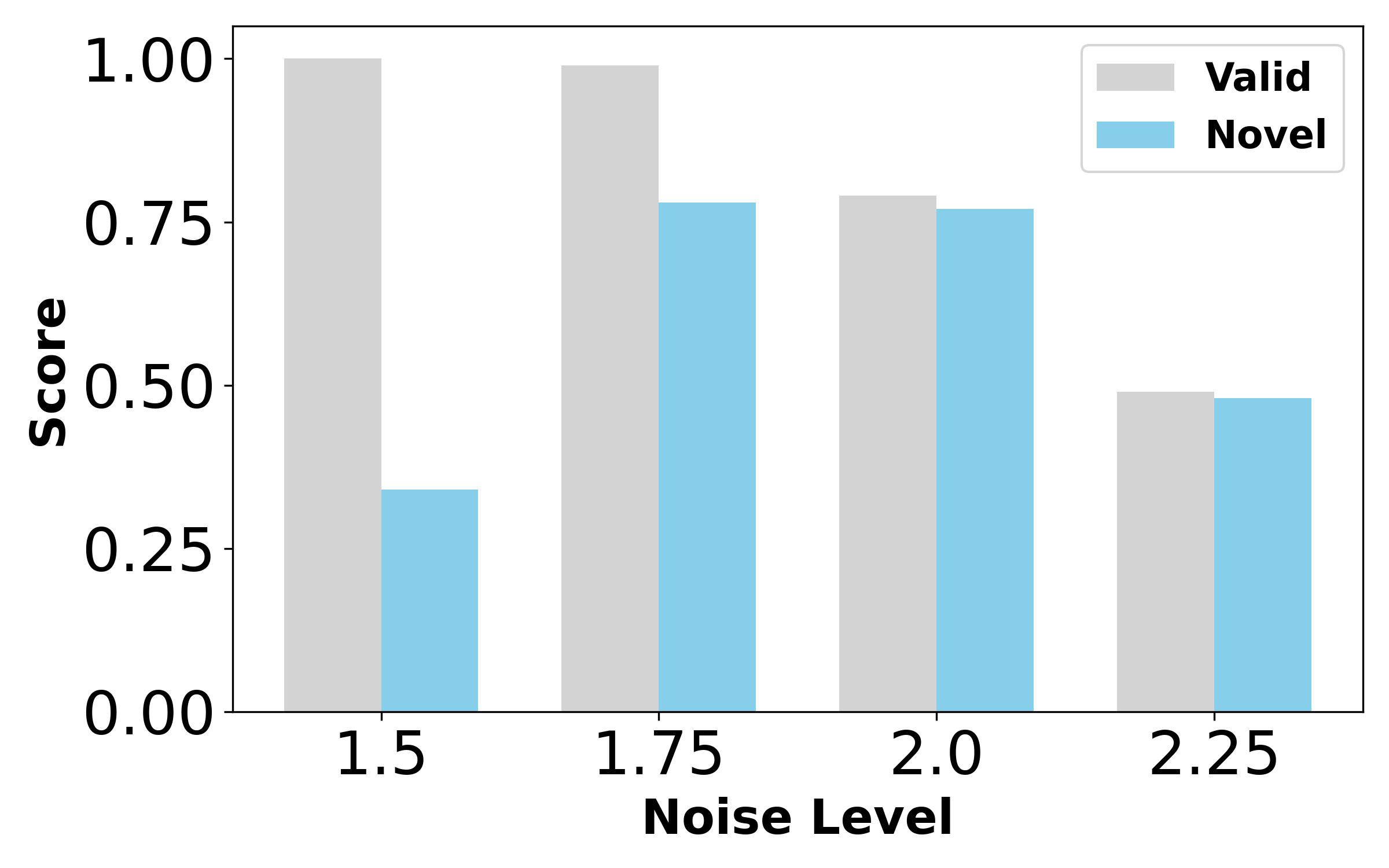}
    \caption{polyBART\textsubscript{large}}
    \label{fig:latent-eg}
\end{subfigure}

\caption{Validity and novelty across noise levels.}
\label{fig:latent-space-merged}
\end{figure}

\section{Metrics}
\label{sec:metrics}
\subsection{FCD} 
FCD is computed using molecular representations extracted from the penultimate layer activations of the ChemNet model. Assuming these activations follow multivariate Gaussian distributions, the mean and covariance is calculated for both generated and input molecules. If the generated distribution \( G \) has mean \( \mu_G \) and covariance \( \Sigma_G \), and the input distribution \( I \) has mean \( \mu_I \) and covariance \( \Sigma_I \), FCD between these two is given by:
\[
\makebox[\columnwidth][c]{%
\resizebox{0.95\columnwidth}{!}{$
\mathrm{FCD}(G, I) = \|\mu_G - \mu_I\|_2^2 + 
\mathrm{Tr}\left(\Sigma_G + \Sigma_I - 2(\Sigma_G \Sigma_I)^{1/2}\right)
$}%
}
\]

\subsection{IntDiv\textsubscript{p}}
IntDiv\textsubscript{p} is computed using \( T(m_1, m_2) \), which denotes the Tanimoto similarity between molecules \( m_1 \) and \( m_2 \). The summation is performed over all pairs of molecules in the set of generations \( G \), and the result is normalized by \( |G|^2 \), the total number of pairwise comparisons. The parameter \( p \) controls the order of the mean. IntDiv\textsubscript{p} is given by:
\[
\makebox[\columnwidth][c]{%
\resizebox{0.9\columnwidth}{!}{$
\mathrm{IntDiv}_p(G) = 1 - \left( \frac{1}{|G|^2} \sum_{m_1, m_2 \in G} T(m_1, m_2)^p \right)^{\frac{1}{p}}
$}%
}
\]

\section{Threshold Selection for SA Scores}
\label{sec:sa_score}

We assess the distribution of SA scores within the experimental T\textsubscript{g} dataset. As shown in Figure~\ref{fig:sa_scores}, majority of the polymers exhibit SA scores of 6 or lower, suggesting that a cutoff of 6 effectively identifies polymers that are sufficiently easy to synthesize. Hence, we adopt this threshold for our computational experiments on chemical structure generation.

\begin{figure}[H]
    \centering
    \includegraphics[width=0.93\columnwidth]{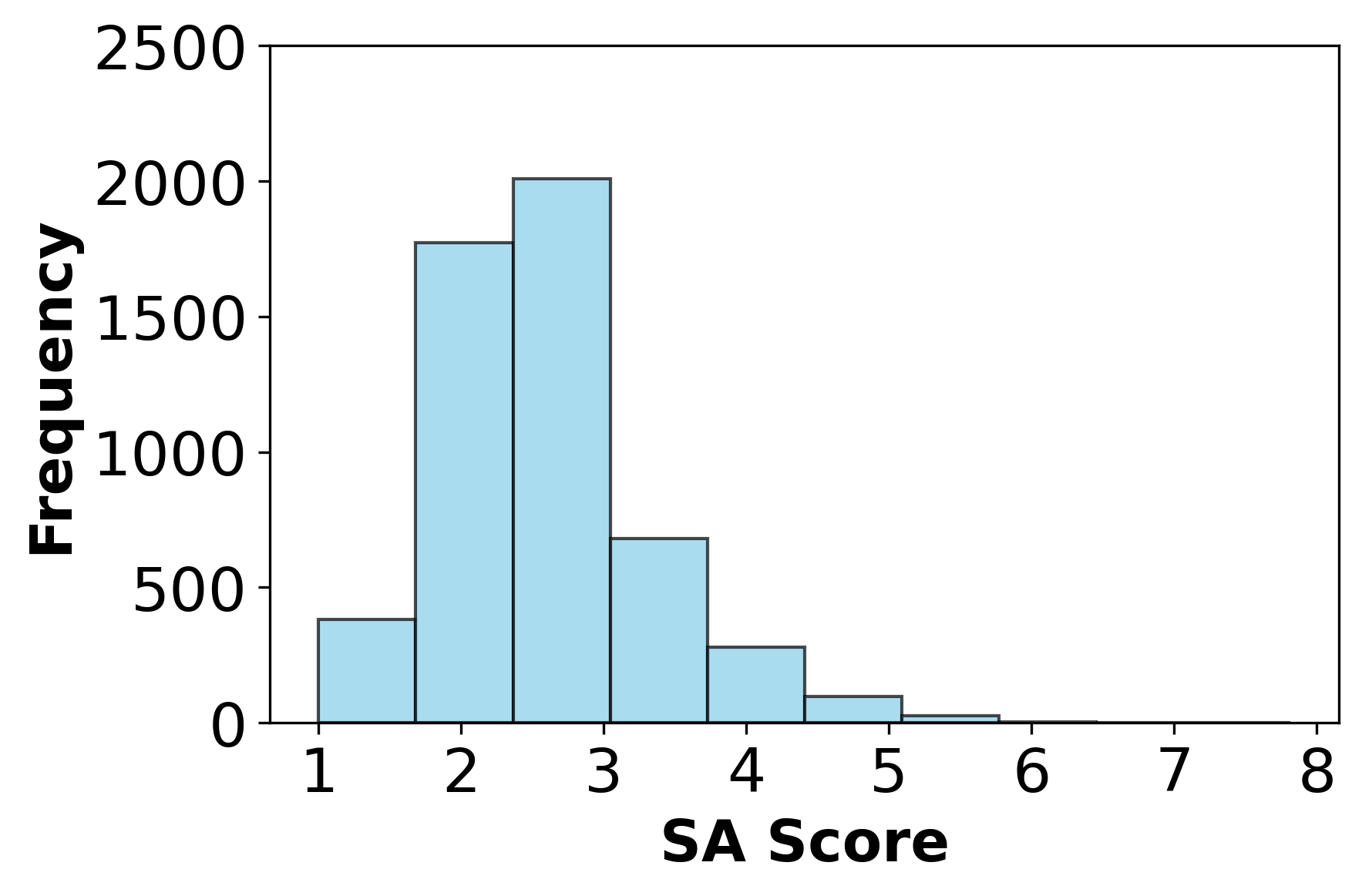}
    \caption{SA score distribution.}
    \label{fig:sa_scores}
\end{figure}

\section{Latent Space Visualization}

\begin{figure*}[t]
\centering
  \includegraphics[width=\textwidth]{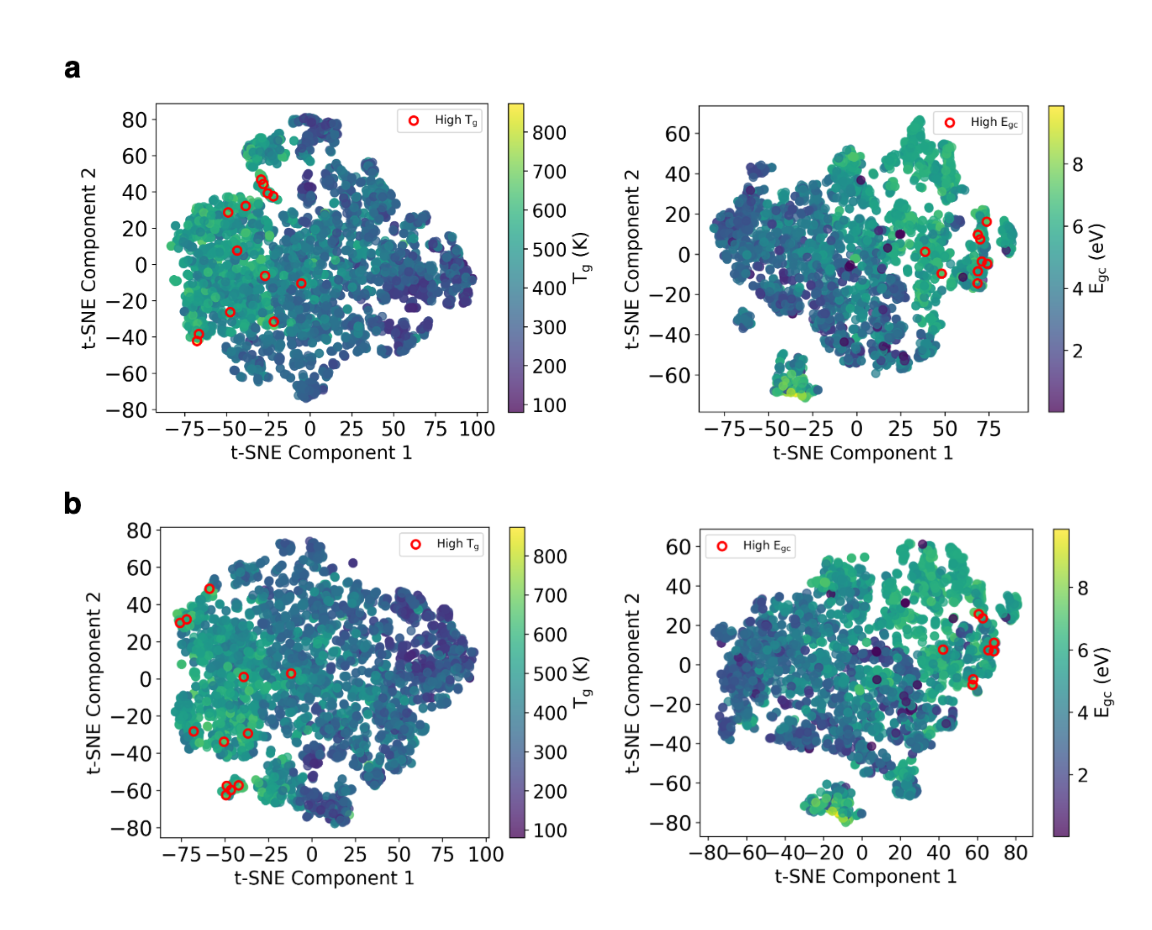}
  \caption{Latent space projections for (a) polyBART\textsubscript{small} and (b) polyBART\textsubscript{large} highlighting polymers exhibiting high T\textsubscript{g} and E\textsubscript{gc} values.}
  \label{fig:latent}
\end{figure*}

Figure~\ref{fig:latent} shows the property-conditioned latent space visualizations for polyBART\textsubscript{small} and polyBART\textsubscript{large}, with each plot highlighting regions in the embedding space associated with polymers exhibiting high T\textsubscript{g} and E\textsubscript{gc} values. The latent representations are obtained by encoding polymer structures with each model's encoder and projecting the resulting embeddings into two dimensions using t-SNE. These visualizations reveal how different models structure the latent space with respect to property-relevant features. A gradient in property values across the latent space suggests that the model has learned a meaningful representation in which polymer embeddings are aligned with their associated properties.
\label{sec:latent_space}

\section{Experimental Procedure}

To synthesize poly(amic acid) (PAA), 3,3$'$-diaminodiphenylmethane (31~mg, 0.16~mmol) is added to a three-necked round-bottom flask, followed by the addition of \textit{N,N}-dimethylacetamide (DMAc, 250~\textmu L) as the solvent. The mixture is stirred until the diamine was completely dissolved. Subsequently, an equimolar amount of 3,3$'$,4,4$'$-benzophenonetetracarboxylic dianhydride (50~mg, 0.16~mmol) is slowly added to the solution under a nitrogen atmosphere. The reaction temperature is maintained between 40$^{\circ}$C and 50$^{\circ}$C, and the mixture is stirred overnight to obtain a homogeneous PAA solution. The resulting solution is slowly poured into deionized water under vigorous stirring, leading to the formation of a white to pale-yellow precipitate. The precipitate is collected by filtration and washed thoroughly with acetone multiple times to remove residual solvent. The purified PAA is air-dried at room temperature and then vacuum-dried at 60$^{\circ}$C until a constant weight is achieved.

Thermal imidization of the obtained PAA is conducted to afford the corresponding polyimide (PI). The dried PAA is placed in a flask under vacuum, and the temperature is gradually increased in a stepwise manner to 150$^{\circ}$C, 200$^{\circ}$C, and finally 300$^{\circ}$C, holding each temperature for approximately 1~hour. After cooling to room temperature under vacuum, the resulting PI (\textasciitilde54~mg, 71\%) is collected for further characterization.

Thermogravimetric analysis (TGA) was performed using a Pyris 1 TGA (PerkinElmer) at a heating rate of 10$^{\circ}$C/min. Differential Scanning Calorimetry (DSC) was carried out on a DSC 3+ STARe system (Mettler Toledo), with both heating and cooling rates set to 10$^{\circ}$C/min.

\begin{figure}[H]
    \centering
    \includegraphics[width=\columnwidth]{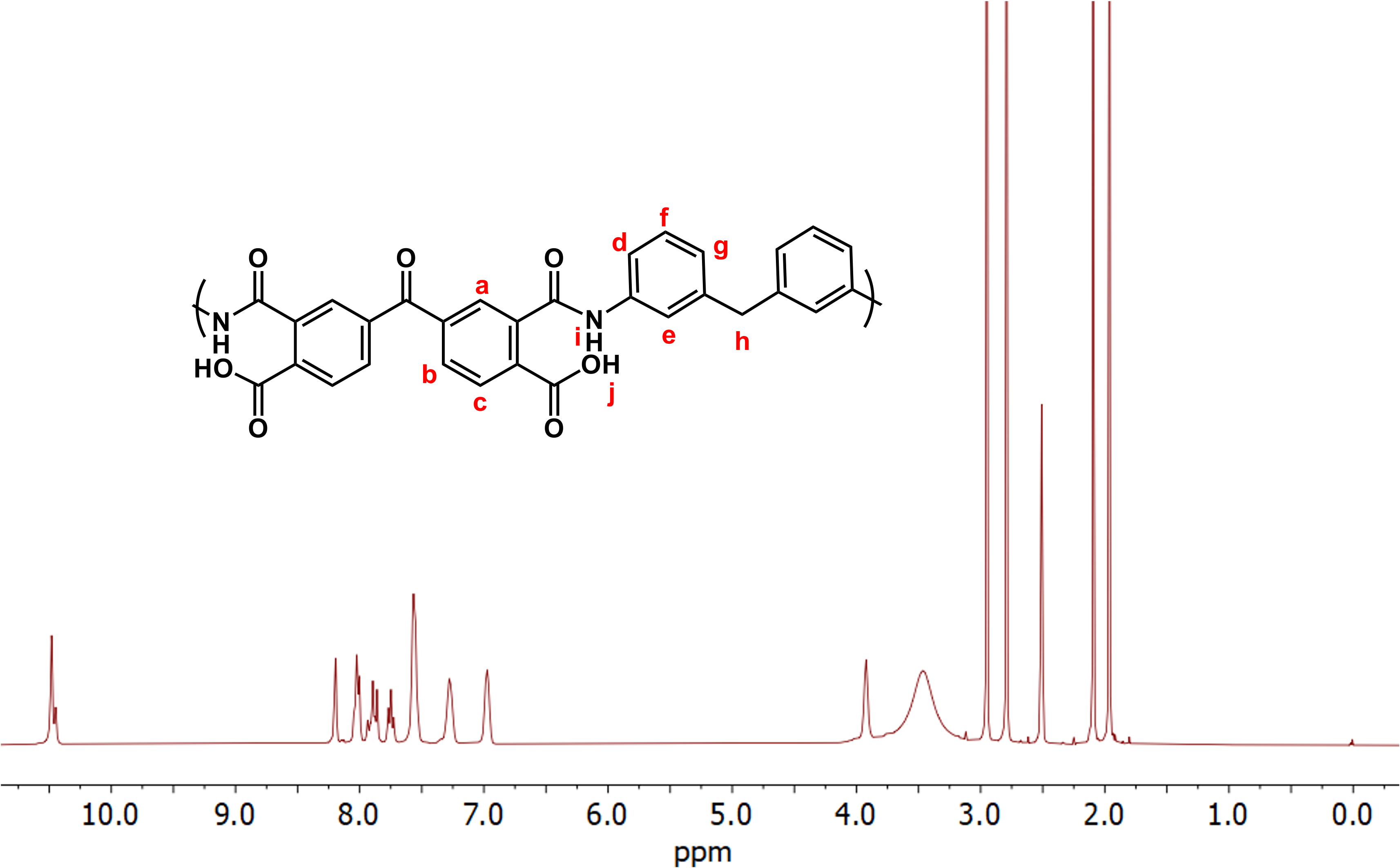}
    \caption{\textsuperscript{1}H NMR spectrum of PAA in d\textsubscript{6}-DMSO.}
    \label{fig:sa_scores}
\end{figure}

\begin{figure}[H]
    \centering
    \includegraphics[width=\columnwidth]{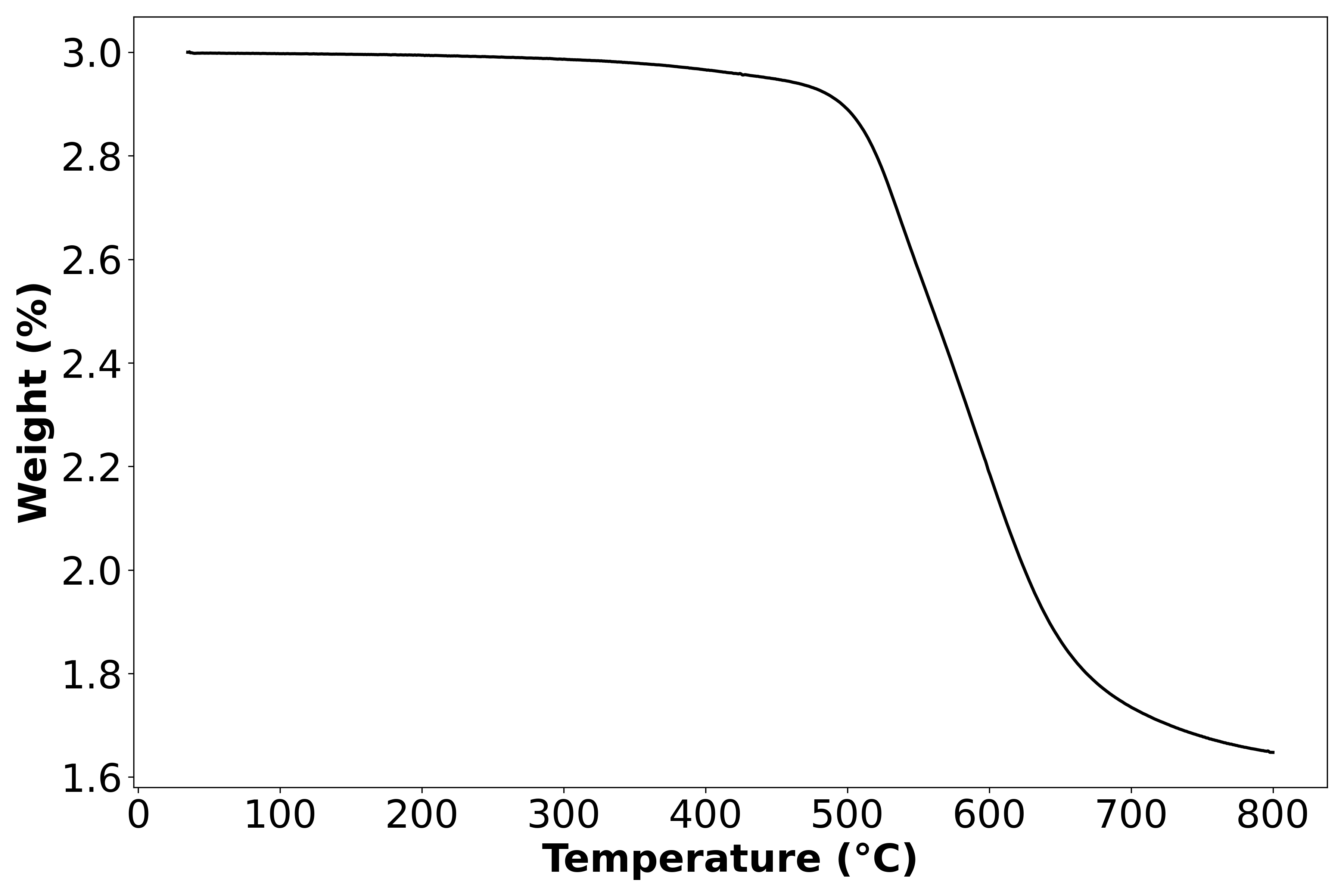}
    \caption{TGA curve of PI.}
    \label{fig:sa_scores}
\end{figure}

\begin{figure}[H]
    \centering
    \includegraphics[width=\columnwidth]{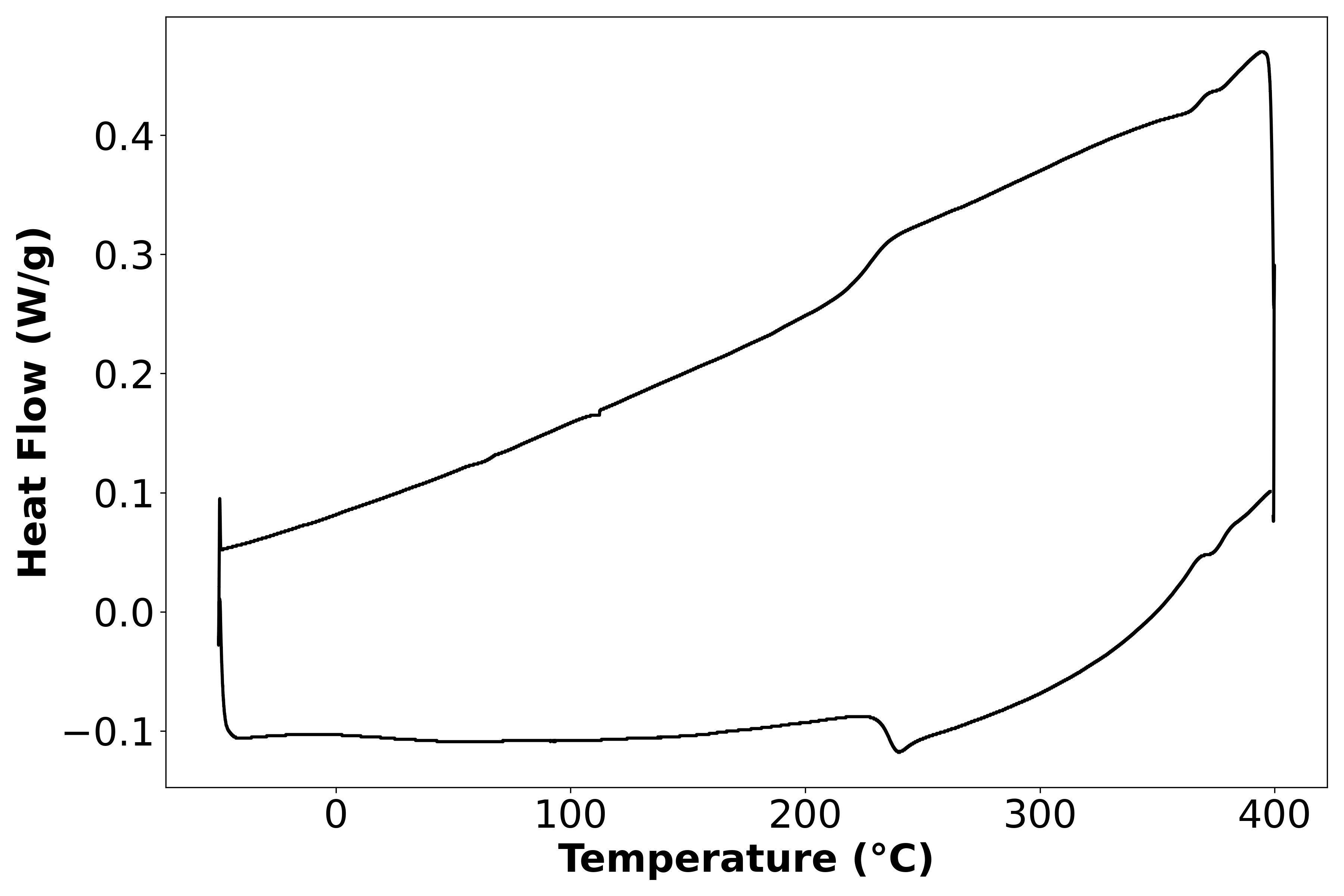}
    \caption{DSA curve of PI.}
    \label{fig:sa_scores}
\end{figure}

\label{sec:synthesis}

\end{document}